\documentclass[12pt]{iopart}

\usepackage[utf8]{inputenc}

\usepackage[english]{babel}

\usepackage{amsmath,amsfonts,amssymb,mathtools}
\usepackage {graphicx}
\graphicspath{{fig/}}
\usepackage{subcaption}

\usepackage{xcolor}

\usepackage{makecell}
\setcellgapes{4pt}

\usepackage{booktabs}
\usepackage{multirow}
\usepackage{graphicx}

\usepackage{cite}
\usepackage{hyperref}
\hypersetup{
	colorlinks=true,        
	linkcolor=blue,          
	citecolor=blue,        
	filecolor=magenta,      
	urlcolor=blue           
}

\usepackage{array}
\usepackage{comment}

\newcommand{\sign}{{\rm sign}}

\usepackage{color} 				
\definecolor{dgray}{rgb}{0.6,0.6,0.6}
\definecolor{dmag}{rgb}{0.6,0.0,0.6}
\definecolor{mbul}{rgb}{0.102, 0.42, 0.102} 
\usepackage[normalem]{ulem} 	

\definecolor{pink}{rgb}{1,0,0.9}

\newcommand{\pd}{{\phantom{\dagger}}}

\begin{document}

\title[]{\bf Broken-axisymmetry state and magnetic state diagram of spin-1 condensate through the prism of quadrupole degrees of freedom} 

\author{M Bulakhov$^{1,2,\dagger}$ , A S Peletminskii$^{1,2,*}$, S V Peletminskii$^1$ and Yu~V~Slyusarenko$^{1,2}$}
\address{$^1$ Akhiezer Institute for Theoretical Physics, National Science Center "Kharkov Institute of Physics and Technology", NAS of Ukraine, 61108 Kharkiv, Ukraine}
\address{$^2$ V.N. Karazin Kharkiv National University, 61022 Kharkiv, Ukraine}
\ead{$^\dagger$ \mailto{bulakh@kipt.kharkov.ua}}
\ead{$^*$ \mailto{aspelet@kipt.kharkov.ua}}




\begin{abstract}
We theoretically study a weakly interacting gas of spin-1 atoms with Bose-Einstein condensate in external magnetic field within the Bogoliubov approach. To this end, in contrast to previous studies, we employ the general Hamiltonian, 
which includes both spin and quadrupole exchange interactions as well as the couplings of the spin and quadrupole moment with the external magnetic field (the linear and quadratic Zeeman terms). The latter is responsible for the emergence of the broken-axisymmetry state. We also reexamine ferromagnetic, quadrupolar, and paramagnetic states employing the proposed Hamiltonian.
For all magnetic states, we find the relevant thermodynamic characteristics such as magnetization, quadrupole moment, thermodynamic potential. We also obtain three-branch excitation spectrum of the broken-axisymmetry state. We show that this state can be prepared at three different regimes of applied magnetic field. 
Finally, we present the magnetic state diagrams for each regime of realizing the broken-axisymmetry state. 
\end{abstract}

\maketitle

\section{Introduction}
Ultracold high-spin atomic gases with Bose-Einstein condensate provide a high-controlled system to examine various magnetic states and corresponding phase transitions \cite{UedaPhysRep,UedaRMP}. The uniqueness of such systems is that they exhibit the simultaneous manifestation of superfluid and magnetic properties. The first theoretical studies \cite{Ohmi1998,Bigelow1998,HoPRL1998,JETP1998} of the so-called spinor condensates were stimulated by the optical trapping of sodium atoms \cite{Stamper1998}, which allows them to be trapped independently of their spin orientations. Therefore, the system is described by the vector order parameter representing a condensate wave function.

Usually, theoretical description of spinor condensates is based on the interaction Hamiltonian with the coupling constants parameterized by the $s$-wave scattering lengths \cite{Ohmi1998,Bigelow1998,HoPRL1998}. In this case, the pairwise many-body Hamiltonian of spin-1 atoms is specified by two scattering lengths $a_{0}$ and $a_{2}$ corresponding to the collisions of two atoms with total spin ${\cal S}=0, 2$. Therefore, it contains two interaction terms: the first one is independent of spin operators, while the second term is bilinear in them. In the presence of the linear and quadratic Zeeman terms, this Hamiltonian produces the ferromagnetic, quadrupolar (polar), paramagnetic (antiferromagnetic), and broken-axisymmetry states \cite{UedaPhysRep,UedaRMP}. In contrast to other magnetic states, the broken-axisymmetry state is only realized for the Hamiltonian with both linear and quadratic Zeeman terms. For the particular regimes of magnetic fields, it was examined theoretically \cite{UedaPRA2007} as well as experimentally \cite{Ketterle1998,Dalibard2012,Chapman2016}.
        
Despite the fact that the scattering-length approximation has proven to be a powerful tool for describing the interaction effects in ultracold gases \cite{Pethick2002,Pitaevskii2003}, it has a number of weak points. In particular, when studying a weakly interacting Bose gas with condensate within the Bogoliubov approach \cite{Bogoliubov1947}, it results in divergence of the ground state energy (or chemical potential) so that some artificial renormalization procedure is needed to remove them in the second order terms in creation and annihilation operators \cite{Pethick2002,Pitaevskii2003} and in higher order terms \cite{Peletminskii2017}. The emergence of divergence is due to the fact that the scattering-length approximation neglects the non-local character of the interparticle interaction, which, nevertheless, is always inherent in it. Moreover, for spinor condensates, this approximation does not reproduce the complete structure of the single-particle excitation spectrum \cite{PhysLettA2020}. The role of non-local interaction has been recently discussed in physics of ultracold Bose \cite{Bulakhov2018,Haas2018} and Fermi \cite{Hara2012,Caballero2013,Simonucci2011} gases.    
Besides that, the applicability of the $s$-wave scattering-length approximation is limited by the choice of the interaction potential. Actually, if the interaction $V\propto r^{-n}$, 
then for $n>3$, this approximation is valid. 
Thus, for atoms with a large dipole moment (see Refs.~\cite{Aikawa_PRL2012, Ferlaino_PRA2022} for Erbium (Er) and \cite{Lu_PRL2011, Lev_PRA2015} for Dysprosium (Dy)) we cannot neglect the dipole-dipole interaction ($V\propto r^{-3}$) and, therefore, directly use the $s$-wave scattering-length approximation. 

In this paper, in contrast to the above studies of spinor condensates, we refuse the scattering-length approximation assuming the interatomic interaction to be of non-local nature. In this case, the corresponding many-body Hamiltonian of spin-1 atoms should be constructed of the spin and quadrupole operators, which are introduced in an elegant way through the prism of the SU(3) algebra on an equal footing \cite{PhysLettA2020}. In particular, three Gell-Mann generators are identified with three components of the spin operator and the remaining five ones, representing the quadrupole degrees of freedom, are organized into the quadrupole matrix. In our previous study \cite{JPhysA2021}, we have already employed the Hamiltonian with quadrupole-quadrupole interaction to examine the ferromagnetic, quadrupolar, and paramagnetic states of a weakly interacting Bose gas with condensate in a magnetic field within the Bogoliubov approach \cite{Bogoliubov1947}. The coupling with external magnetic field was provided by means of a spin operator (the so-called linear Zeeman shift).
Meanwhile, it is clear that the quadrupole degrees of freedom should also be coupled with the external field. Therefore, here we study all possible magnetic states using the most complete Hamiltonian with both couplings as well as quadrupole-quadrupole and spin-spin exchange interactions. It is worth stressing that exactly the magnetic anisotropy induced by the coupling of the quadrupole moment with the external field gives rise to the fourth broken-axisymmetry state, theoretically predicted earlier for a more simple model \cite{UedaPRA2007}. We also determine and analyze the additional constraints on magnetic fields at which the broken-axisymmetry state can be realized. This allows us to present the magnetic state diagrams for all regimes of the broken-axisymmetry state as well as to investigate the possible phase transitions between all magnetic states. 

\section{Hamiltonian with quadrupole degrees of freedom}

Let us briefly remind the main steps in theoretical description of interacting gas of spin-1 atoms. 
For low-energy collisions, the pairwise many-body Hamiltonian of spin-1 atoms is specified by two coupling constants expressed in terms of the scattering lengths $a_{0}$ and $a_{2}$. The latter correspond to the collisions of two atoms with total spins ${\cal S}= 0$ and ${\cal S}=2$, respectively. The scattering with ${\cal S}=1$ is forbidden, since in $s$-state of relative motion the unit total spin is ruled out by the requirement for the wave function to be symmetric under exchange of two atoms. Therefore, the resulting interaction Hamiltonian contains two terms – the first one is independent of spin-1 operators, while the second term is bilinear in them \cite{Ohmi1998,Bigelow1998,HoPRL1998}. This description assumes the point-like (or local) interaction between atoms.

However, for non-local interaction, the mentioned form of the Hamiltonian is insufficient for a correct description of the system due to the fact that the scattering-length approximation fails and the scattering of two atoms with total spin ${\cal S}=1$ is now possible. In this case, the interaction Hamiltonian should be constructed of the spin and quadrupole operators, which are introduced through the prism of the SU(3) algebra on an equal footing \cite{PhysLettA2020,JPhysA2021}. In particular, eight Gell-Mann generators $\lambda^{a}$ ($a=1,\dots 8$) split into the three spin component operators $S^{i}=(S^{x}=\lambda^{7},S^{y}=-\lambda^{5},S^{z}=\lambda^{2})$ in the Cartesian basis, $(S^{i})_{kl}=-i\varepsilon_{ikl}$,  and five quadrupole operators $q^{b}=(-\lambda^{1},-\lambda^{3},-\lambda^{4},-\lambda^{6},\lambda^{8})$, see 
\ref{app:Gell} and \ref{ap:sol} 
for details. The latter, according to Eq.~(\ref{eq:AAntiCom}), are expressed in terms of the spin components operators in the following manner:
\begin{gather*}
\lambda^{1}=-\{S^{x},S^{y}\}, \quad \lambda^{3}=(S^{y})^{2}-(S^{x})^{2}, \quad \lambda^{4}=-\{S^{x},S^{z}\}, 
\\ 
\lambda^{6}=-\{S^{y},S^{z}\}, \quad \lambda^{8}=\sqrt{3}(S^{z})^{2}-\frac{2}{\sqrt{3}}I,
\end{gather*}
where $\{a,b\}=ab+ba$ denotes the anticommutator and $I$ is the identity operator. Five components of $q^{b}$ specify the symmetric quadrupole matrix (or nematic tensor) ${\cal Q}^{ik}=S^{i}S^{k}+S^{k}S^{i}-(4/3)\delta^{ik}$ (see. e.g., \cite{Hamley2012,Corboz2018}),
\begin{equation} 
\label{eq:QuadMatr}
{\cal Q}=\left(
  \begin{array}{ccc}
    -\lambda^{3}-\lambda^{8}/\sqrt{3} & -\lambda^{1} & -\lambda^{4} \\ 
    -\lambda^{1} & \lambda^{3}-\lambda^{8}/\sqrt{3} & -\lambda^{6} \\
    -\lambda^{4} & -\lambda^{6} & 2\lambda^{8}/\sqrt{3} \\
  \end{array}
\right).
\end{equation}
The expectation value of the quadrupole matrix describes the anisotropy of spin fluctuations. Thus, just as an electric charge induces all electric moments, the particle spin generates all magnetic moments. 

The grand canonical Hamiltonian (operator that enters the Gibbs statistical operator corresponding to the grand canonical ensemble) of the system can be written in the following form:
\begin{equation}\label{eq:TotalHam}
{\cal H}={\cal H}_{0}+{V}_{U}+{V}_{J}+{V}_{K},
\end{equation}
where
\begin{equation}\label{eq:FreeHam}
{\cal H}_{0}=\sum_{\bf p}a^{\dagger}_{{\bf p}\alpha}\left[(\varepsilon_{\bf p}-\mu)\delta_{\alpha\beta}-h^{i}S^{i}_{\alpha\beta}-
\frac{1}{2}
b^{i}{\cal Q}^{ik}_{\alpha\beta}b^{k}\right]a_{{\bf p}\beta}.
\end{equation}
Here $\varepsilon_{{\bf p}}={{\bf p}^{2}/2m}$ is the kinetic energy of the atom, $\mu$ is the chemical potential, and $a^{\dagger}_{{\bf p}\alpha}$ ($a_{{\bf p}\alpha}$) is the creation (annihilation) operator of bosonic atom in a state with momentum ${\bf p}$ and spin projection $\alpha$. Two other terms containing $h^{i}$ and $b^{i}$ describe the coupling of spin and quadrupole moment with the external field, respectively (see the discussion below).
The isotropic interaction Hamiltonian is specified by the spin-independent term $V_{U}$ as well as the terms $V_{J}$ and $V_{K}$ describing the spin-exchange and quadrupole-exchange interactions, respectively,
\begin{gather} 
V_{U}=\frac1{2\mathcal{V}}\sum_{{\bf p}_{1}\ldots{\bf p}_{4}}U({\bf p}_{1}-{\bf p}_{3})a^{\dagger}_{{\bf p}_{1}\alpha}a^{\dagger}_{{\bf p}_{2}\beta}a_{{\bf p}_{3}\alpha}a_{{\bf p}_{4}\beta}\,\delta_{{\bf p}_{1}+{\bf p}_{2},\,{\bf p}_{3}+{\bf p}_{4}}, \nonumber \\
V_{J}=\frac1{2\mathcal{V}}\sum_{{\bf p}_{1}\ldots{\bf p}_{4}}J({\bf p}_{1}-{\bf p}_{3})a^{\dagger}_{{\bf p}_{1}\alpha}a^{\dagger}_{{\bf p}_{2}\beta}S^{i}_{\alpha \gamma}S^{i}_{\beta\delta}a_{{\bf p}_{3}\gamma}a_{{\bf p}_{4}\delta}\,\delta_{{\bf p}_{1}+{\bf p}_{2},\,{\bf p}_{3}+{\bf p}_{4}}, \nonumber \\
V_{K}=\frac1{4{\cal V}}\sum_{{\bf p}_{1}\dots{\bf p}_{4}}K({\bf p}_{1}-{\bf p}_{3})a^{\dagger}_{{\bf p}_{1}\alpha}a^{\dagger}_{{\bf p}_{2}\beta}{\cal Q}^{ik}_{\alpha\gamma}{\cal Q}^{ki}_{\beta\delta}a_{{\bf p}_{3}\gamma}a_{{\bf p}_{2}\delta}\delta_{{\bf p}_{1}+{\bf p}_{2},{\bf p}_{3}+{\bf p}_{4}}. \label{eq:InterHam}
\end{gather}
Noting that ${\cal Q}^{ik}_{\alpha\gamma}{\cal Q}_{\beta\delta}^{ki}=2q_{\alpha\gamma}^{b}q_{\beta\delta}^{b}$, one obtains \cite{PhysLettA2020,JPhysA2021}
\begin{equation} \label{eq:InterHamK}
V_{K}=\frac1{2{\cal V}}\sum_{{\bf p}_{1}\dots{\bf p}_{4}}K({\bf p}_{1}-{\bf p}_{3})a^{\dagger}_{{\bf p}_{1}\alpha}a^{\dagger}_{{\bf p}_{2}\beta}q^{b}_{\alpha\gamma}q^{b}_{\beta\delta}a_{{\bf p}_{3}\gamma}a_{{\bf p}_{2}\delta}\delta_{{\bf p}_{1}+{\bf p}_{2},{\bf p}_{3}+{\bf p}_{4}}.
\end{equation}
Since 
\begin{equation}\label{eq:Ident}
\frac{1}{2}q^{b}_{\alpha\gamma}q^{b}_{\beta\delta}=S_{\alpha\sigma}^{i}S_{\beta\rho}^{i}S_{\sigma\gamma}^{k}S_{\rho\delta}^{k}+\frac{1}{2}S_{\alpha\gamma}^{i}S_{\beta\delta}^{i}-\frac{4}{3}\delta_{\alpha\gamma}\delta_{\beta\delta},
\end{equation}
the presence of quadrupole degrees of freedom in the interaction Hamiltonian is equivalent to the fact that it contains both bilinear and biquadratic terms in spin operators. It is evident that the Hamiltonian given by Eqs.~(\ref{eq:InterHam}) and (\ref{eq:InterHamK}) has SU(2) symmetry like the bilinear Heisenberg-type Hamiltonian with $s$-wave scattering length parameterization employed in Refs.~\cite{Ohmi1998,HoPRL1998,UedaPRA2007} to study the magnetic properties of spin-1 Bose gas with condensate.
However, the difference is that the Hamiltonian under consideration is more general: it includes the additional quadrupole degrees of freedom 
both in its free and interaction parts. 
For quantum gases, such a structure makes sense if the interatomic interaction is characterized by the finite range. 
Therefore, the Hamiltonian is determined by three Fourier transforms of the corresponding interaction energies. In addition, for $J({\bf p})=K({\bf p})$, the interaction Hamiltonian becomes SU(3) invariant. The similar Hamiltonian with non-local interaction and bilinear term in the spin operators was used to analyze the magnetic states of spin-$S$ condensate with only one component of the order parameter be occupied \cite{JETP1998,PhysicaA2007}.   

Generally speaking, the explicit form of functions $U({\bf p})$, $J({\bf p})$, and $K({\bf p})$ should be found from the Coloumb interaction for a given interatomic species that represents a very difficult and unsolved problem. However, their constant values $U(0)$, $J(0)$, and $K(0)$ can be related to $s$-wave scattering lengths. Within the scattering-length approximation, we can represent the interaction as $V=\sum g_F P_F$, where $P_F$ is a projection operator on the state with total spin $F=0,1,2$ and $g_{F}$ is a corresponding coupling constant. Since the $s$-wave scattering of two identical spin-1 atoms with total spin $F=1$ is ruled out by the requirement for the wave function to be symmetric under the exchange of atoms, the projection operator $P_1=0$ (for details, see Ref.~\cite{PhysLettA2020})). 
On the other hand, assuming interaction to be of contact type, according to Eqs.~(4) and (5), we get
\begin{equation*} 
	V=\frac1{2\mathcal{V}}\sum_{{\bf p}_{1}\ldots{\bf p}_{4}}a^{\dagger}_{{\bf p}_{1}\alpha}a^{\dagger}_{{\bf p}_{2}\beta}
	\left[
		U(0)\delta_{\alpha\gamma}\delta_{\beta\delta}
		+
		J(0)S^{i}_{\alpha \gamma}S^{i}_{\beta\delta}
		+
		K(0)q^{b}_{\alpha \gamma}q^{b}_{\beta\delta}
	\right]
	a_{{\bf p}_{3}\gamma}a_{{\bf p}_{4}\delta}\,\delta_{{\bf p}_{1}+{\bf p}_{2},\,{\bf p}_{3}+{\bf p}_{4}}
	.
\end{equation*}
Next, substituting here the identity given by Eq.~(\ref{eq:Ident})
and employing the definition of the projection operator \cite{PhysLettA2020},
\begin{equation*}
	(P_1)_{\alpha\beta\gamma\delta}
	\equiv
	\delta_{\alpha\gamma}\delta_{\beta\delta}
	-
	\frac12
	\left(
		S_{\alpha\gamma}^{i}S_{\beta\delta}^{i}
		+
		S_{\alpha\sigma}^{i}S_{\beta\rho}^{i}S_{\sigma\gamma}^{k}S_{\rho\delta}^{k}
	\right)
	,\quad
	P_1=
	0,
\end{equation*}
we obtain
\begin{multline}
\label{eq:local}
	V=
	\frac1{2\mathcal{V}}\sum_{{\bf p}_{1}\ldots{\bf p}_{4}}a^{\dagger}_{{\bf p}_{1}\alpha}a^{\dagger}_{{\bf p}_{2}\beta}
	\left[
	\left(U(0)+\frac43K(0)\right)\delta_{\alpha\gamma}\delta_{\beta\delta}
	+
	\bigg(J(0)-K(0)\bigg)S^{i}_{\alpha \gamma}S^{i}_{\beta\delta}
	\right]
	a_{{\bf p}_{3}\gamma}a_{{\bf p}_{4}\delta}\,
	\times\\
	\delta_{{\bf p}_{1}+{\bf p}_{2},\,{\bf p}_{3}+{\bf p}_{4}}
	.
\end{multline}
Finally, comparing this result with the above definition of interaction expressed through the projection operators, we conclude that the interaction in spin-1 system can be specified by two coupling constants, 
\begin{equation*}
\tilde{v}\equiv U(0)+\frac43 K(0)=\frac{g_{0}+2g_{2}}{3},  \quad  
\tilde{c}\equiv J(0)-K(0)=\frac{g_{2}-g_{0}}{3},
\end{equation*}
where
$$
g_{0}=\frac{4\pi\hbar^{2}}{m}a_{0}, \quad g_{2}=\frac{4\pi\hbar^{2}}{m}a_{2}.
$$ 
Here $a_0$ and $a_2$ are the $s$-wave scattering lengths corresponding to the scattering of two atoms with total spin $F=0$ and $F=2$, respectively. The notions $\tilde{v}$ and $\tilde{c}$ are in full agreement with those used below in Eqs.~(\ref{eq:Interparam}). 

The role of $K(0)$ can be essential for long-range atomic interaction. In particular, the atoms with large magnetic moments are subject to dipole-dipole interaction and the quantities $U(0)$, $J(0)$, and $K(0)$ can be of the same order of magnitude (for details, see Section \ref{Summary}).

Let us return to the Hamiltonian ${\cal H}_{0}$.
Assuming that vectors $h^{i}$ and $b^{i}$ are directed along $z$-axis, $h^{i}=(0,0,h)$ and $b^{i}=(0,0,\sqrt{\chi})$, one obtains  
\begin{equation}\label{eq:FreeHam_z}
{\cal H}_{0}=\sum_{\bf p}a^{\dagger}_{{\bf p}\alpha}\left[(\varepsilon_{\bf p}-\tilde{\mu})\delta_{\alpha\beta}-hS^{z}_{\alpha\beta}-\chi(S^{z})^{2}_{\alpha\beta}\right]a_{{\bf p}\beta},
\end{equation}
where
\begin{equation} \label{eq:ChemPot}
\tilde{\mu}=\mu-\frac{2}{3}\chi.
\end{equation}
Some comments should be made regarding $h$ and $\chi$. The terms $hS^{z}$ and $\chi(S^{z})^{2}$ are associated with the linear and quadratic Zeeman energies, respectively. Thus, $h=g\mu_{B}B$, where $g$ is the Land\'{e} hyperfine factor, $\mu_{B}$ is the Bohr magneton, and $B$ is an external magnetic field directed along $z$-axis. The parameter $\chi$ includes the contribution from the external field $B$,
$$
\chi_{B}=\frac{(g\mu_{B}B)^{2}}{\Delta E},
$$
where $\Delta E=E_{m}-E_{g}$ is the hyperfine energy splitting given by the difference between
the intermediate $E_{m}$ and ground $E_{g}$ energies. Another contribution to $\chi$ is experimentally possible due to a microwave \cite{Gerbier2006,Leslie2009} or light \cite{Santos2007} field. Therefore, the coefficients $h$ and $\chi$ can be varied independently \cite{Ketterle1998}. It is worth noting that the quadratic Zeeman term naturally appears in the Hamiltonian (see Eq.~(\ref{eq:FreeHam})) when we attempt to include the coupling of quadrupole degrees of freedom (or quadrupole moment) with an external field. Therefore, we can affirm that the quadratic Zeeman term determines the magnetic anisotropy of ultracold quantum gases.

\section{Ferromagnetic, quadrupolar, and paramagnetic states of spin-1 BEC within the Bogoliubov model}

Here we apply the grand canonical Hamiltonian given by Eqs.~(\ref{eq:TotalHam}), (\ref{eq:InterHam}), (\ref{eq:InterHamK}), and (\ref{eq:FreeHam_z}) to examine the magnetic states that can emerge in a weakly interacting gas of spin-1 atoms with Bose-Einstein condensate. To this end, we employ the Bogoliubov model \cite{Bogoliubov1947,Bogol_Intr} for a weakly interacting Bose gas, which treats the creation and annihilation operators with zero momentum as $c$-numbers. Following it, one should replace $a_{0\alpha}\to\sqrt{\cal V}\Psi_{\alpha}$ and $a^{\dagger}_{0\alpha}\to\sqrt{\cal V}\Psi^{*}_{\alpha}$ in all relevant operators of physical quantities, where $\Psi_{\alpha}$ is the order parameter or the condensate wave function. Performing this replacement in the Hamiltonian, it is sufficient to restrict ourselves by the $c$-number terms and those, which are quadratic in creation and annihilation operators of atoms with nonzero momentum but to neglect the higher order terms, relevant when describing the interaction between quasiparticles. Therefore, the resulting grand canonical Hamiltonian reads,
\begin{equation}\label{eq:BogolHam}
{\cal H}(\Psi)\approx {\cal H}^{(0)}(\Psi)+{\cal H}^{(2)}(\Psi),  
\end{equation}
where ${\cal H}^{(0)}(\Psi)$ is its $c$-number part specifying the ground state,  
\begin{multline}\label{eq:cNumbHam}
\frac1{\mathcal{V}}{\cal H}^{(0)}(\Psi)=\frac{U(0)} 2(\Psi^{*}\Psi)^{2}+\frac{J(0)}2(\Psi^{*}S^{i}\Psi)^{2}+
\frac{K(0)}2(\Psi^{*}q^{b}\Psi)^{2} 
\\
-h(\Psi^{*}S^{z}\Psi)-\chi(\Psi^{*}({S^{z}})^{2}\Psi)-\tilde{\mu}(\Psi^{*}\Psi). 
\end{multline}
Here and below, one should understand the brackets as a summation over repeated indices, e.g., $(\Psi^{*}\Psi)\equiv \Psi^{*}_{\alpha}\Psi^{\pd}_{\alpha}$ or $(\Psi^* q^{b} \Psi)^{2} \equiv (\Psi^{*}_\alpha q^{b}_{\alpha\beta}\Psi_{\beta}) (\Psi^{*}_\gamma q^{b}_{\gamma\delta}\Psi_{\delta})$.
The part of the Hamiltonian that is quadratic in creation and annihilation operators with nonzero momentum, ${\cal H}^{(2)}(\Psi)$, has the form
\begin{gather}
{\cal H}^{(2)}(\Psi)=\sum_{{\bf p}\neq0}a^{\dagger}_{{\bf p}\alpha}\left[(\varepsilon_{\bf p}-\tilde{\mu})\delta_{\alpha\beta}- hS^{z}_{\alpha\beta}-\chi(S^{z})^{2}_{\alpha\beta}\right]a_{{\bf p}\beta} \label{eq:QuadrHam} \\ 
+
U(0)\sum_{{\bf p}\neq0}(\Psi^{*}\Psi)(a^{\dagger}_{\bf p}a_{\bf p})+\frac12\sum_{{\bf p}\neq 0}U({\bf p})\left[(a^{\dagger}_{\bf p}\Psi )(\Psi^{*}a_{\bf p})+(a^{\dagger}_{\bf p}\Psi)(a^{\dagger}_{-{\bf p}}\Psi )+{\rm h.c.}\right] \nonumber \\
+J(0)\sum_{{\bf p}\neq 0}(\Psi^{*}S^{i}\Psi)(a^{\dagger}_{{\bf p}}S^{i}a_{\bf p}) 
	+\frac12
	\sum_{{\bf p}\neq 0}J({\bf p})\left[(a^{\dagger}_{\bf p}S^{i}\Psi)(\Psi^{*}S^{i}a_{\bf p})+(a^{\dagger}_{\bf p}S^{i}\Psi)(a^{\dagger}_{-{\bf p}}S^{i}\Psi)+{\rm h.c.}\right] \nonumber \\
+ K(0)\sum_{{\bf p}\neq 0}
	(\Psi^{*}q^{b}\Psi)(a^{\dagger}_{{\bf p}}q^{b}a_{\bf p}) 
	+
	\frac12
	\sum_{{\bf p}\neq 0}K({\bf p})\left[(a^{\dagger}_{\bf p}q^{b}\Psi)(\Psi^{*}q^{b}a_{\bf p})+(a^{\dagger}_{\bf p}q^{b}\Psi)(a^{\dagger}_{-{\bf p}}q^{b}\Psi)+{\rm h.c.}\right]. \nonumber
\end{gather}
After the corresponding diagonalization with respect to creation and annihilation operators, ${\cal H}^{(2)}(\Psi)$ determines the spectrum of single-particle excitations for a given ground state.
The Gibbs statistical operator for the Bogoliubov model reads,
\begin{eqnarray*}\label{eq:GibbsOper}
w(\Psi)\approx\exp\left(\beta[\Omega(\Psi)-{\cal H}(\Psi)]\right),
\end{eqnarray*}
where $\beta=1/T$ is the reciprocal temperature and $\Omega$ is the Gibbs thermodynamic potential, which is found from the normalization condition ${\rm Tr}\,w=1$,
\begin{eqnarray*}\label{eq:GrandPot}
\Omega(\Psi)={\cal H}^{(0)}(\Psi)-\frac{1}{\beta}\ln{\rm Tr}\left[ \exp(-\beta{\cal H}^{(2)}(\Psi))\right].
\end{eqnarray*}
Next, it is convenient to introduce the density of thermodynamic potential $\varpi(\Psi)=\Omega(\Psi)/{\cal V}$. Up to a sign, it coincides with pressure $P$, $\varpi=-P$. The thermodynamic stability of the system requires the pressure to be positive ($\varpi$ must be negative, $\varpi<0$). In the main approximation of the model (neglect of ${\cal H}^{2}(\Psi)$), the density of thermodynamic potential has the form
\begin{equation}\label{eq:GrandPotDens}
\varpi\approx \frac{U(0)}2(\Psi^{*}\Psi)^{2}+\frac{J(0)}2(\Psi^{*}S^{i}\Psi)^{2}+\frac{K(0)}2(\Psi^{*}q^{b}\Psi)^{2}-h(\Psi^{*}S^{z}\Psi)-\chi(\Psi^{*}(S^{z})^{2}\Psi)-\tilde{\mu}(\Psi^{*}\Psi).
\end{equation}
Then, in agreement with the standard Bogoliubov model, the ground-state structure is found by minimization of $\varpi$ with respect to the condensate wave function, $\partial\varpi/\partial\Psi_{\alpha}^{*}=0$. This yields 
\begin{equation}
\label{eq:MinCond}
\tilde{\mu}\zeta_{\alpha}+hS_{\alpha\beta}^{z}\zeta_{\beta}+\chi(S^{z})^{2}_{\alpha\beta}\zeta_{\beta}-n_{0}U(0)\zeta_{\alpha}-n_{0}J(0)(\zeta^{*}S^{i}\zeta)S^{i}_{\alpha\beta}\zeta_{\beta}-n_{0}K(0)(\zeta^{*}q^{b}\zeta)q^{b}_{\alpha\beta}\zeta_{\beta}=0,
\end{equation}
where $n_{0}$ is the condensate density and $\zeta_{\alpha}$ is a normalized vector defined as
\begin{equation} \label{eq:NormSpinor}
\Psi_{\alpha}=\sqrt{n_{0}}\zeta_{\alpha}, \quad \zeta_{\alpha}^{\phantom{*}}\zeta_{\alpha}^{*}=1.
\end{equation}

The system of coupled equations (\ref{eq:MinCond}) has four different solutions (see 
\ref{ap:sol}). Three of them are isomorphic to those obtained in Ref.~\cite{JPhysA2021}, where only the linear Zeeman term was considered in the Hamiltonian. This means that the state vectors $\zeta_{\alpha}$ preserve their form and by setting the terms generated by $\chi$ 
equal to zero, the physical quantities such as the chemical potential and the density of thermodynamic potential become those of Ref.~\cite{JPhysA2021}. These three solutions correspond to ferromagnetic (F), quadrupolar (Q), and paramagnetic (P) states. Their physical characteristics are summarized in the Table~\ref{tab:3states}.
\begin{table}[htb!]
\centering
\setlength{\tabcolsep}{0.4mm}
\makegapedcells
\begin{tabular}{@{}cccccc@{}}
\toprule 
\multirow{2}{*}{\bf State} & 
\multicolumn{2}{c}{\bf State vector}
&
\multirow{2}{*}{\parbox{6.62em}{\bf Chemical\newline potential, $\tilde{\mu}$}} & \multirow{2}{*}{\parbox{6.7em}{\bf Density of\newline potential, $\varpi/n_0$}} & \multirow{2}{*}{\parbox{5em}{\bf\centering Magneti\-zation, $\langle S^{z}\rangle/n_0$}}
\\\cmidrule{2-3} 
    &{\bf$\boldsymbol{\zeta}$ (Cartesian)}&{\bf $\boldsymbol{\zeta}^{\prime}$ (canonical)}&&&\\[1ex] 
\midrule
F$_{\pm}$ & 
$\displaystyle
    \frac{1}{\sqrt{2}}(\mp 1,-i,0)^{T}
$
&

\parbox{7em}{\centering
$\displaystyle(1,0,0)^{T}$
\\
$\displaystyle(0,0,1)^{T}$
}
& 
$\displaystyle{v+c-\chi\mp h}$ 
& 
$\displaystyle \frac{v+c}{2}-\chi\mp h-\tilde{\mu}$
&
$\displaystyle \pm 1$
\\
Q & 
$(0,0,1)^{T}$ 
& 

$\displaystyle{(0,1,0)^{T}}$
&
$\displaystyle v$
& 
$\displaystyle\frac{v}{2}-\tilde{\mu}$
&
$\displaystyle 0$
\\
P & 
$\displaystyle-(a,ib,0)^{T}$ 
&
$\displaystyle\left(\frac{a+b}{\sqrt{2}},0,\frac{b-a}{\sqrt{2}}\right)^{T}$
& 
$\displaystyle v-\chi$
&
$\displaystyle \frac{v}{2}-\frac{h^{2}}{2c}-\chi-\tilde{\mu}$
&
$\displaystyle \frac{h}{c}$
\\
\bottomrule
\end{tabular}
\nomakegapedcells
\small
\flushleft
\begin{tabular*}{\linewidth}{l@{}ll}
*&$a=\left(e^{i\phi_{+}}\sqrt{1+h/c}+e^{i\phi_{-}}\sqrt{1-h/c}\right)/2$ and &$b=\left(e^{i\phi_{+}}\sqrt{1+h/c}-e^{i\phi_{-}}\sqrt{1-h/c}\right)/2$.\\ 
&$\phi_{+}$ and $\phi_{-}$ are the arbitrary phases.
\end{tabular*}
\caption{Thermodynamic characteristics of ferromagnetic (F), quadrupolar (Q), and paramagnetic (P) states of spin-1 condensate.}
\label{tab:3states}
\end{table}
In particular, the chemical potential is found by solving Eq.~(\ref{eq:MinCond}) (for details, see 
\ref{ap:sol}). The density of thermodynamic potential $\varpi$ for each magnetic state is obtained from Eq.~(\ref{eq:GrandPotDens}) by substituting Eq.~(\ref{eq:NormSpinor}) with corresponding state vector $\zeta_{\alpha}$. The magnetization is obtained from 
\begin{equation}
    \label{eq:MagVec}
    \langle S^{i}\rangle=\Psi_{\alpha}^{*}S^{i}_{\alpha\beta}\Psi_{\beta},
\end{equation}
where $\Psi_{\alpha}$ is determined by Eqs.~(\ref{eq:NormSpinor}) (for ferromagnetic and paramagnetic states only, $\langle S^{z}\rangle$ is different from zero). 
As a result of calculations, all physical characteristics are expressed in terms of the following quantities:
\begin{equation}\label{eq:Interparam}
v=n_{0}\left(U(0)+\frac43K(0)\right), \quad c=n_{0}\bigg(J(0)-K(0)\bigg).
\end{equation}
Therefore, parameter $c$ represents the measure of the SU(3) symmetry breaking for pairwise interaction. Note that for ferromagnetic state ($F_{+}$) with positive magnetization, the state vector differs by a sign from that used in our previous work \cite{JPhysA2021}.
In fact, both state vectors are equivalent due to the phase invariance of Eq.~\eqref{eq:MinCond}. Here we change the sign to ensure the correct transformation between the representations of spin matrices and state vectors in Cartesian and canonical bases.
In addition to the three obtained solutions of Eq.~(\ref{eq:MinCond}), there is only one more fourth solution, which we discuss in the next section.

\section{Broken-axisymmetry state} 

The fourth solution of Eq.~(\ref{eq:MinCond}) emerges due to the term $\chi(S^{z})^{2}_{\alpha\beta}$ describing the coupling of the quadrupole moment with external field. Following \cite{UedaPRA2007,Dalibard2012}, we call the state corresponding to this solution as broken-axisymmetry (BA) state. The reason for this name will be clear below.
Note that there are no other solutions of Eq.~(\ref{eq:MinCond}). 

The corresponding solution of Eq.~(\ref{eq:MinCond}) in canonical basis reads (see 
\ref{ap:sol}),
\begin{equation}\label{eq:BAChem}
\boldsymbol{\zeta}^\prime=(\zeta_{+},\zeta_{0},\zeta_{-})^{T}, \quad \tilde{\mu}=\frac{h^{2}-\chi^{2}}{2\chi}+v+c,  
\end{equation}
where
\begin{equation}\label{eq:BAState}
\zeta_{\pm}=\left[\frac{(h\mp\chi)^{2}}{8c\chi^{3}}\left(h^{2}-\chi^{2}+2c\chi\right)\right]^{1/2}e^{i\phi_{\pm}}, 
\quad \zeta_{0}=\left[\frac{\chi^{2}-h^{2}}{4c\chi^{3}}(h^{2}+\chi^{2}+2c\chi)\right]^{1/2}e^{i\phi_{0}}.
\end{equation}
The state vector in Cartesian basis, $\boldsymbol{\zeta}=(\zeta_{x},\zeta_{y},\zeta_{z})^{T}$, is related to that in canonical basis, $\boldsymbol{\zeta}^\prime$, by the unitary transformation, $\boldsymbol{\zeta}=R^{\dagger}\boldsymbol{\zeta}^\prime$, (see Eq.~(\ref{eq:bastrans})). Thus, for the components of $\boldsymbol{\zeta}$, we have 
\begin{equation*}\label{eq:BAStateVect}
\zeta_{x}=\frac1{\sqrt{2}}(\zeta_{-}-\zeta_{+}), \quad \zeta_{y}=-\frac{i}{\sqrt{2}}(\zeta_{+}+\zeta_{-}), \quad \zeta_{z}=\zeta_0.
\end{equation*}
According to Eq.~(\ref{eq:ForPhase}), the phases \(\phi_{+}\), \(\phi_{-}\), and \(\phi_{0}\) are related by the following equation:
\begin{equation}\label{eq:BAPhases}
\phi_{+}+\phi_{-}-2\phi_{0}=\arg\left[\sign(h^2-\chi^2)\right].
\end{equation}
Therefore, all the state vector components are occupied in the BA-state. By the way, spin domain formation in spin-1 condensates was extensively studied both theoretically and experimentally \cite{Ketterle1998,Isoshima1999,Stamper1999,Matuszewski2008,De2014,Dalibard2019}.  
 
We must also ensure that the radicands in Eqs.~(\ref{eq:BAState}) are positive. Thus, we have the following three cases:
\begin{subequations}
\begin{alignat}{1}
\label{ineq:cxl0xlh}
c\chi<0,\quad &
\chi^{2}-h^{2}\leq 0, \quad h^{2}-\chi^{2}+2c\chi\leq 0, \quad h^{2}+\chi^{2}+2c\chi\geq 0
, \\
\label{ineq:cxl0xgh}
c\chi<0,\quad &
\chi^{2}-h^{2}\geq 0, \quad h^{2}-\chi^{2}+2c\chi\leq 0, \quad h^{2}+\chi^{2}+2c\chi\leq 0
, \\
\label{ineq:cxg0xgh}
c\chi>0,
\quad &
\chi^{2}-h^{2}\geq 0, \quad h^{2}-\chi^{2}+2c\chi\geq 0,
\quad h^{2}+\chi^{2}+2c\chi\geq 0.
\end{alignat}
\label{eq:ineq}
\end{subequations}
Each system of inequalities (\ref{eq:ineq}) determines the specific regime of realizing the broken-axisymmetry state. All of them are examined below employing the general Hamiltonian under consideration. 
Note, the regime (\ref{ineq:cxg0xgh}) was studied previously in Refs.~\cite{Ketterle1998,UedaPRA2007} on the basis of Hamiltonian with spin-independent and bilinear spin-exchange interactions parameterized by the $s$-wave scattering lengths. For regime (\ref{ineq:cxl0xgh}), the phase transition from BA-state to paramagnetic one was examined experimentally \cite{Dalibard2012}.

Having defined the state vector and corresponding constraints determined by (\ref{eq:ineq}), we can obtain the main thermodynamic characteristics of BA-state such as magnetization, density of thermodynamic potential (or pressure), expectation value of the quadrupole matrix, and single-particle excitation spectrum. 

We begin with the components of magnetization vector. According to its general definition (\ref{eq:MagVec}), one obtains   
\begin{gather*}
\langle S^{x} \rangle=\sqrt{2}n_{0}{\rm Re}\left[\zeta_{0}^{*}(\zeta_{-}-\zeta_{+})\right], \quad \langle S^{y} \rangle=\sqrt{2}n_{0}{\rm Im}\left[\zeta_{0}^{*}(\zeta_{-}+\zeta_{+})\right], \\
\langle S^{z}\rangle=n_{0}(|\zeta_{+}|^{2}-|\zeta_{-}|^{2}).
\end{gather*}
Substituting here the explicit expressions for $\zeta_{\pm}$ and $\zeta_{0}$ from Eq.~(\ref{eq:BAState}), one obtains
\begin{gather*}
\langle S^{x} \rangle=\frac{n_{0}\sqrt{(\chi^2-h^2)((h^2+2c\chi)^2-\chi^4)}}{4|c\chi|\chi^{2}}\left(|\chi+h|\cos{(\phi_{-}-\phi_{0})}-|\chi-h|\cos{(\phi_{+}-\phi_{0})} \right), \\
\langle S^{y}\rangle=\frac{n_{0}\sqrt{(\chi^2-h^2)((h^2+2c\chi)^2-\chi^4)}}{4|c\chi|\chi^{2}}\left(|\chi+h|\sin{(\phi_{-}-\phi_{0})+|\chi-h|\sin{(\phi_{+}-\phi_{0})}} \right), \\
\langle S^{z}\rangle=-n_{0}h\frac{h^{2}-\chi^{2}+2c\chi}{2c\chi^{2}}. 
\end{gather*}
While $\langle S^{z}\rangle$ is the same for all regimes of magnetic fields (see inequalities (\ref{eq:ineq})), the components $\langle S^{x}\rangle$ and $\langle S^{y}\rangle$ depend 
on the sign of $\chi^{2}-h^{2}$. Therefore, removing moduli and using Eq.~(\ref{eq:BAPhases}) to fix the phases as $\phi_{
+}+\phi_{-}-2\phi_{0}=0$ for $\chi^{2}-h^{2}\leq 0$ and $\phi_{
+}+\phi_{-}-2\phi_{0}=\pi$ for $\chi^{2}-h^{2}\geq 0$, we come to the final expressions for $\langle S^{x}\rangle$ and $\langle S^{y}\rangle$. The results are summarized in the Table~\ref{tab:MagComp}.  

\begin{table}[htb!]
    \makegapedcells
    \centering
    \begin{tabular}{@{}cccc@{}}
        \toprule
        \parbox{7.3em}{\bf Magnetization components} & {\bf General formula} & $\chi^2-h^2>0$ & $\chi^2-h^2<0$\\
        \midrule
        $\langle S^{x}\rangle$ & $\sqrt{2}n_0{\rm Re}\,\zeta_0^*\left(\zeta_{-}-\zeta_{+}\right)$ & $\displaystyle\langle S^{\perp}\rangle\sin\frac{\Delta \phi}{2}$ & $\displaystyle\langle S^{\perp}\rangle\cos\frac{\Delta \phi}{2}\sign(\chi h)$\\
        $\langle S^{y}\rangle$ & $\sqrt{2}n_0{\rm Im}\,\zeta_0^*\left(\zeta_{-}+\zeta_{+}\right)$ & $\displaystyle\langle S^{\perp}\rangle\cos\frac{\Delta \phi}{2}$ & $\displaystyle-\langle S^{\perp}\rangle\sin\frac{\Delta \phi}{2}\sign(\chi h)$\\
        $\langle S^{z}\rangle$ & $n_0\left(|\zeta_{+}|^2-|\zeta_{-}|^2\right)$ & \multicolumn{2}{c}{$\displaystyle-n_{0}h
\,\frac{h^{2}-\chi^{2}+2c\chi}{2c\chi^2}$}\\
        \bottomrule
    \end{tabular}
    \nomakegapedcells
    \flushleft
    \begin{tabular*}{0.5\linewidth}{l@{}ll}
    *&$\displaystyle\langle S^{\perp}\rangle\equiv
    \sqrt{\langle S^{x}\rangle^{2}+\langle S^{y}\rangle^{2}}=\frac{n_0}{2|c|\chi^2}\sqrt{(\chi^2-h^2)((h^2+2c\chi)^2-\chi^4)},$  
    &$\Delta\phi=\phi_{+}-\phi_{-}$
    \end{tabular*}
    \caption{Components of the magnetization vector for magnetic field regimes $\chi^{2}-h^{2}\geq 0$ and $\chi^{2}-h^{2}\leq 0$, see~Eqs.~(\ref{eq:ineq}).}
    \label{tab:MagComp}
\end{table}
\noindent
The fact that $\left<S^{x}\right>\neq 0$ and $\left<S^{y}\right>\neq 0$ indicates that the magnetization vector is not parallel to the applied magnetic field directed along $z$-axis, i.e., it has the perpendicular component $\langle S^{\perp}\rangle$. 
The angle $\theta$ between the magnetization vector and direction of the field is determined from the following equation:
\begin{equation*}
\tan\theta=\frac{\langle S^{\perp}\rangle}{\langle S^{z}\rangle}=-\frac{\sign(c)}{h}\frac{\sqrt{(\chi^2-h^2)((h^2+2c\chi)^2-\chi^4)}}{h^{2}-\chi^{2}+2c\chi}.    
\end{equation*}
Hence, we can conclude that both axial and more general rotational symmetries are broken. It is useful to obtain the modulus of magnetization vector,
\begin{equation*}
\label{eq:magnet_lf}
    M
    \equiv
    |\langle {\bf S}\rangle|
    =
    \frac{n_{0}}{2|c\chi|}
    \bigg[
    4c^2\chi^2
    -
    (\chi^{2}-h^{2})^2
    \bigg]^{1/2}.
\end{equation*}

The density of thermodynamic potential $\varpi$ as a function of $n_0$ and $\mu$ is obtained by substituting Eq.~(\ref{eq:NormSpinor}) with corresponding state vector  into Eq.~\eqref{eq:GrandPotDens}. As a result, $\varpi$ can be expressed in terms of the above magnetization,
\begin{equation} \label{eq:ThermPotLZ}
    \varpi(n_0)
    =
    n_0
    \left(
    \frac{v}{2} 
    -
   \frac{c}{2}
    \left(\frac{M}{n_0}\right)^2
    +
    \frac{h^{2}-\chi^{2}+2c\chi}{2\chi}
    -
    \tilde{\mu}
    \right).
\end{equation}
For the stability of the system, it is necessary that the density of thermodynamic potential be negative, $\varpi<0$ (the pressure must be positive). 

The expectation value of the quadrupole operator, according to Eq.~(\ref{eq:QuadMatr}), is found to be
\begin{gather} \label{eq:Q}
    \langle {\cal Q}\rangle\equiv\Psi_{\alpha}^{*}{\cal Q}_{\alpha\beta}\Psi_{\beta}=
    \nonumber\\ 
        n_0
        \left(
    \begin{array}{>{\displaystyle}c>{\displaystyle}c>{\displaystyle}c}
        \displaystyle
        |\zeta_z|^2-\frac13-\gamma_{\rm BA}\cos \Delta\phi 
        &
        \gamma_{\rm BA}\sin \Delta\phi
        &
        -\frac{h}{|\chi|}\langle S^{x}\rangle 
        \\[2ex]
        \gamma_{\rm BA}\sin \Delta\phi
        &\displaystyle
        |\zeta_z|^2-\frac13+\gamma_{\rm BA}\cos \Delta\phi
        &
        -\frac{h}{|\chi|}\langle S^{y}\rangle
        \\[2ex]
        -\frac{h}{|\chi|}\langle S^{x}\rangle
        &
        -\frac{h}{|\chi|}\langle S^{y}\rangle  &
        -2|\zeta_z|^2+\frac23
    \end{array}
    \right),
\end{gather}
where 
\begin{equation*}
    \gamma_{\rm BA}=\frac{|(h^{2}-\chi^{2})(h^2-\chi^2+2c\chi)|}{4|c\chi^3|}, \quad \Delta\phi=\phi_{+}-\phi_{-}.
\end{equation*}
For example, the fact that $\langle{\cal Q}^{xz}\rangle\neq \langle{\cal Q}^{yz}\rangle$ and $\langle{\cal Q}^{xx}\rangle\neq \langle{\cal Q}^{yy}\rangle$ demonstrates that the axial and rotational symmetries about $z$-axis, respectively,  
are broken as we found out above. 
We can see that the broken-axisymmetry state is characterized by the significant magnetic anisotropy introducing by the quadrupole operator ${\cal Q}^{ik}$ (see Eqs.~(\ref{eq:FreeHam}) and (\ref{eq:FreeHam_z})). Note that there is a coordinate system, where the quadrupole operator, $\langle {\cal Q}\rangle$, has a diagonal form and does not depend on phases $\phi_{+}$ and $\phi_{-}$.

The excitation spectrum is obtained by reducing the total Hamiltonian determined by Eqs.~(\ref{eq:BogolHam})--(\ref{eq:QuadrHam}) to the diagonal form:
$$
U{\cal H}(\Psi)U^{\dagger}=\sum_{{\bf p},\alpha}\omega_{{\bf p}\alpha}\,a^{\dagger}_{{\bf p}\alpha}a_{{\bf p}\alpha}+{\cal E}_{0},
$$
where the unitary operator $U$ ($UU^{\dagger}=1$) mixes the creation and annihilation operators \cite{AkhPel,PhysLettA2020}. The quantity ${\cal E}_{0}$ determines the ground-state thermodynamic potential in the Gibbs statistical operator including the contribution from the quadratic terms in creation and annihilation operators. The diagonalization of the Hamiltonian becomes more simple in the coordinate system, where the quadrupole matrix given by Eq.~(\ref{eq:Q}) has a diagonal form. This can be achieved by the prior orthogonal transformation, which also makes the matrix $\langle{\cal Q}\rangle$ being phase-independent and the Hamiltonian depending on phases only through trivial multiplier $e^{\pm i(\phi_{+}+\phi_{-})}$. Therefore, in contrast to Ref.~\cite{UedaPhysRep,UedaPRA2007}, where the phases are removed artificially, we can do this naturally by applying the orthogonal transformation.

Then, the standard diagonalization procedure \cite{Bogol_Intr} leads to three branches of the single-particle excitation spectrum. In the linear order in $\varepsilon_{\bf p}$, the gap branch of the spectrum reads,    
\begin{equation}
    \label{eq:spec1}
    \omega_{{\bf p}}^{2}\approx
    \omega^2_{0}-\left(4\frac{h^2+c\chi}{\chi}+G\right)\varepsilon_{\bf p},
\end{equation}
where
\begin{equation}\label{eq:gap}
    \omega^2_{0}=\frac{\left(3h^2+2c\chi-\chi^2\right)\left(h^2+\chi^2+2c\chi\right)}{\chi^2}.
\end{equation}
Since the interaction is assumed to be of non-local type, the excitation energies are also specified by the first derivatives of the interaction amplitudes with respect to energy at zero momentum. These derivatives contribute to $G$,   
\begin{equation*}
G=
\frac{(h^2-\chi^2)}{c}
    \left[1+\frac{h^2(h^2+2c\chi+\chi^2)}{\chi^2(3h^2+2c\chi-\chi^2)}\right]n_0 J^{\prime}(0)+ 
    2\left[4\frac{h^2+c\chi}{\chi}+\frac{(h^2-\chi^2)}{c}\right]n_0 K^{\prime}(0),
\end{equation*}
where 
\begin{equation*}
J'(0)\equiv \left.\frac{\partial J({\bf p})}{\partial\varepsilon_{\bf p}}\right|_{{\bf p}=0}, \quad K'(0)\equiv \left.\frac{\partial K({\bf p})}{\partial\varepsilon_{\bf p}}\right|_{{\bf p}=0}.
\end{equation*}
Note that the gap in Eq.~(\ref{eq:gap}) is due to external magnetic field. 
Within the standard Bogoliubov or Gross-Pitaevskii approaches with corresponding determination of the chemical potential the energy gap also appears for molecular condensates \cite{Radzihovsky2008,Poluektov2014}. In this case the Hamiltonian involves the processes associated with conversion of two atoms into bound state and vice versa.

Two other branches of the single-particle excitations are gapless,  
\begin{equation}
\label{eq:spec2}
    \frac{\omega_{{\bf p}\pm}^2}{\varepsilon_{{\bf p}}}\approx
    v+c+\frac{h^2-\chi^2}{2\chi}
    \pm 
 \sqrt{\left(v+c-\frac{h^2-\chi^2}{2\chi}+D\right)^2-
                (v+c)
        \frac{
            2h^2
            (h^2-\chi^2)^2
        }{c\chi^2(3 h^2+2c\chi-\chi^2)},
    }   
\end{equation}
where
\begin{equation*}
D=
n_0
\frac{(\chi^2-h^2)\left(h^2+2c\chi-\chi^2\right) }{4 c \chi^2}
   \left(
        \frac{h^2+2c\chi+\chi^2}{3 h^2+2c\chi-\chi^2}J^{\prime}(0)
        +
        K^{\prime}(0)
    \right).
\end{equation*}
For $K(0)=0$, $G=0$, and $D=0$, the energies of the single-particle excitations coincide with those obtained in Ref.~\cite{UedaPRA2007}.

\section{Stability conditions and magnetic state diagrams}

The necessary condition for thermodynamic stability of each studied magnetic state is determined by the negative value of the corresponding density of thermodynamic potential after the substituting respective chemical potential, see Table~\ref{tab:3states} and Eq.~(\ref{eq:ThermPotLZ}). This requirement, equivalent to the positiveness of the pressure ($\varpi=-P$),  imposes certain constraints on the interaction parameters, which are summarized in Table~\ref{tab:StabCond}. Moreover, we also indicate the additional inequalities, which guarantee the radicands of the state vector components to be positive for paramagnetic and broken-axisymmetry states. 
\begin{table}[htb!]
    \centering 
    \makegapedcells
    \begin{tabular}{cc@{}}
    \toprule
    {\bf State} & \parbox{12em}{\bf Stability conditions}\\
    \midrule
        F$_{\pm}$ &  $v+c>0$\\
        Q &  $v>0$\\
        P &  $\displaystyle v+\frac{h^2}{c}>0,\quad |h|\leq |c|$\\
        BA&  $\displaystyle v+\frac{c}{n_0^2}M^2(n_0)>0$,\quad Eqs.~\eqref{eq:ineq}\\
    \bottomrule
    \end{tabular}
    \caption{Stability conditions for ferromagnetic (F), quadrupolar (Q), paramagnetic (P), and broken-axisymmetry (BA) states.}
    \label{tab:StabCond}
\end{table}

The next step for constructing a state diagram is to determine the curves that separate the regions of the possible existence of the studied magnetic states. These curves, called separatrices, can be found by equating chemical potentials for the respective pair of states. 
In addition, by comparing the densities of thermodynamic potentials, we can determine which of the two given states is thermodynamically favourable and, therefore, ascertain  the possibility of a phase transition between the corresponding magnetic states on a given separatrix. In Table~\ref{tab:Separ},
we present the equations of all separatrices as well as the results of comparing the densities of thermodynamic potentials for different signs of the interaction parameter $c$ (see Eq.~(\ref{eq:Interparam})). The latter determines the magnetic properties of the system and characterizes the measure of the SU(3) symmetry breaking of pairwise interaction. The phase transition across the separatix is allowed for a state characterized by the greater absolute value of thermodynamic potential. 
\begin{table}[htb!]
    \centering
    \makegapedcells
    \begin{tabular}{cccccc@{}}
    \toprule
    \multirow{2}{*}{\#} &   \multirow{2}{*}{\bf Pair of states}  &  \multirow{2}{*}{\bf Separatrix} & 
    \multicolumn{2}{c}{\bf Preference}
    & \multirow{2}{*}{\parbox{3em}{\bf PhT order}}
    \\\cmidrule{4-5} 
    &&& $c>0$ & $c<0$&\\
    \midrule
    I$_{\pm}$       &   F$_{\pm}$\,/\,Q         &  $h=\pm c\mp\chi$                & F$_{\pm}$  & Q    & $\rm 1^{st}$\\
    II      &   Q\,/\,BA        &  $h=\sqrt{\chi^2-2c\chi}$  & BA & Q    & $\rm 2^{nd}$\\ 
    III     &   P\,/\,BA        &  $h=\sqrt{-\chi^2-2c\chi}$ & P  & BA   & $\rm 2^{nd}$\\
    IV$_{\pm}$      &   F$_{\pm}$\,/\,P         &  $h=\pm c$                     & F$_{\pm}$  & P    & $\rm 2^{nd}$\\
    V$_{\pm}$       &   F$_{\pm}$\,/\,BA        &  $h=\mp\chi$                 & F$_{\pm}$  & BA   & $\rm 2^{nd}$\\
    VI      &   Q\,/\,P       &  $\chi=0$                    & P  & Q    & $\rm 1^{st}$\\
    \bottomrule
    \end{tabular}
    \caption{Separatrix equations, thermodynamic preference of magnetic states for different signs of $c$, and phase transition (PhT) order.}
    \label{tab:Separ}
\end{table}

\begin{figure}[htb]
\begin{minipage}[t]{0.49\linewidth}
    \centering
    \includegraphics[width=48mm]{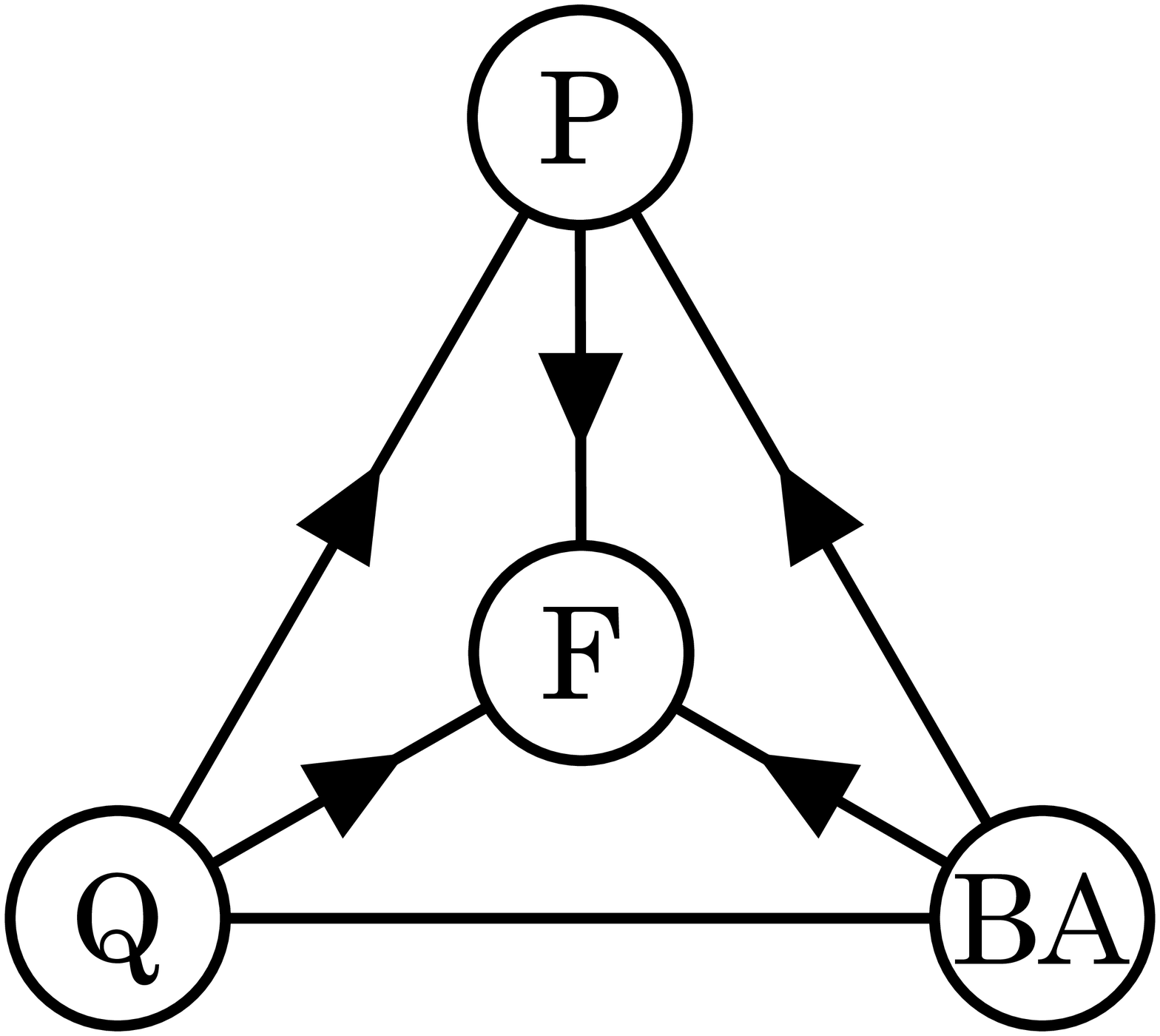}
    \subcaption{$c>0$}
    \label{fig:graph_c>0}
\end{minipage}   
\hfill
\begin{minipage}[t]{0.49\linewidth}
    \centering
    \includegraphics[width=48mm]{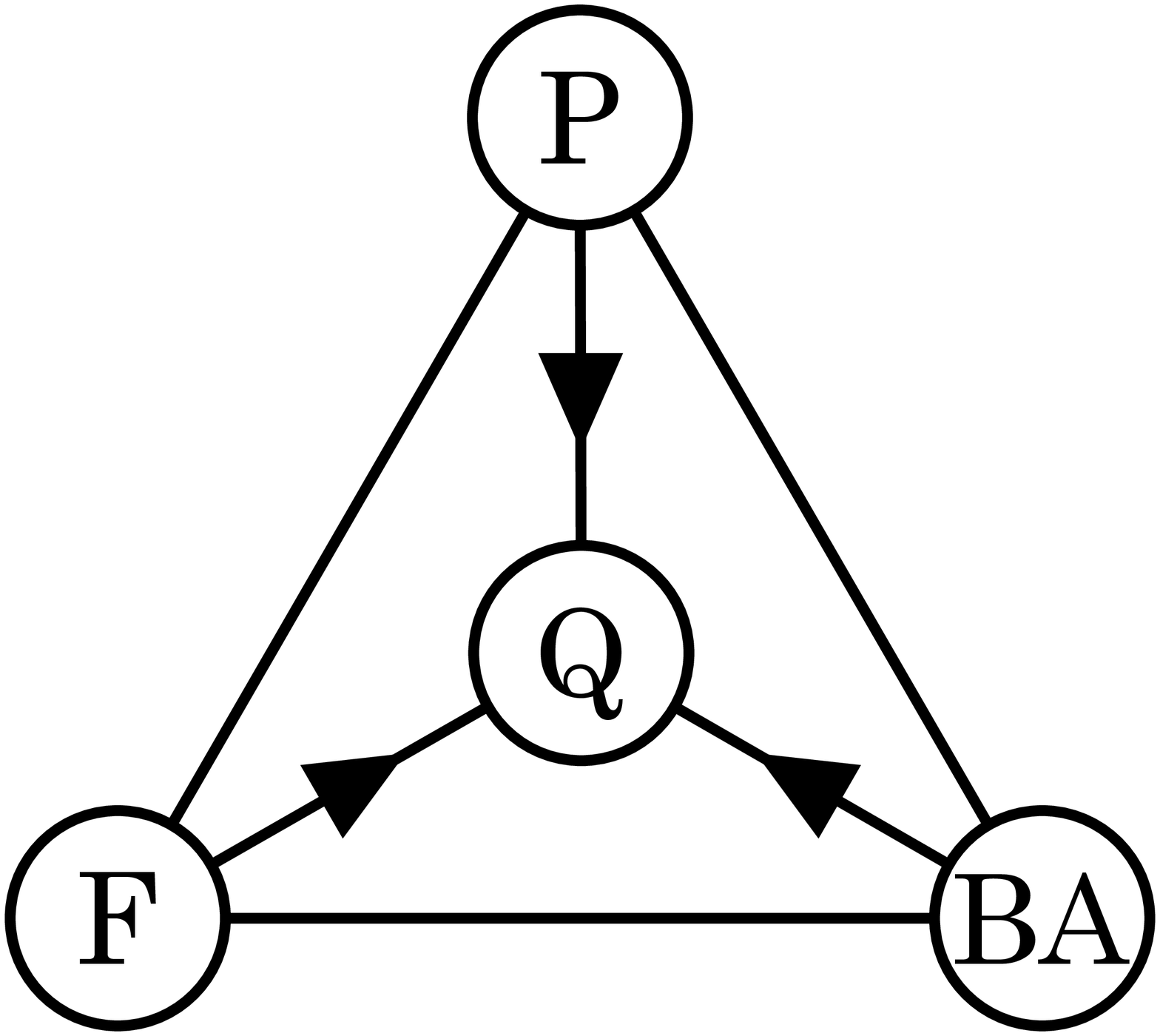}
    \subcaption{$c<0$ }
    \label{fig:graph_c<0}
\end{minipage} 
\caption{Graphs representing the directions and successions of phase transitions for different signs of interaction parameter $c$.}
\label{fig:graph}
\end{figure}

It is  useful to visualize the results of Table~\ref{tab:Separ} in the form of two graphs shown in Fig.~\ref{fig:graph}. 
Since the system cannot leave a more thermodynamically favourable state on a separatrix, it is convenient to introduce the notion of the direction of phase transition between the magnetic states.
Therefore, the graphs have two kinds of edges: those with arrows indicate the direction of phase transition on a separatrix, and those without them indicate the possibility of 
phase transition in both directions on it. The two-directed phase transition can undergo because some separatrices coincide with conditions given by $|h|\leq |c|$ and Eqs.~(\ref{eq:ineq}), see also Table~\ref{tab:StabCond}. 
The graphs show that for some magnetic states there is only one phase transition (e.g., from paramagnetic to ferromagnetic state Fig.~\ref{fig:graph_c>0}) or the transition is forbidden at all (e.g., from ferromagnetic (quadrupolar) to any other state Fig.~\ref{fig:graph_c>0} (Fig.~\ref{fig:graph_c<0})) under the change of external fields $h$ and $\chi$. Due to this fact, depending on the initial state in which the system is prepared, the magnetic state diagram can exhibit either a relatively poor or sufficiently rich variety of phase transitions, even a "hysteresis".

\begin{figure}[htb!]
    \centering
    \begin{minipage}[t]{0.49\linewidth}
    \includegraphics[width=77mm]{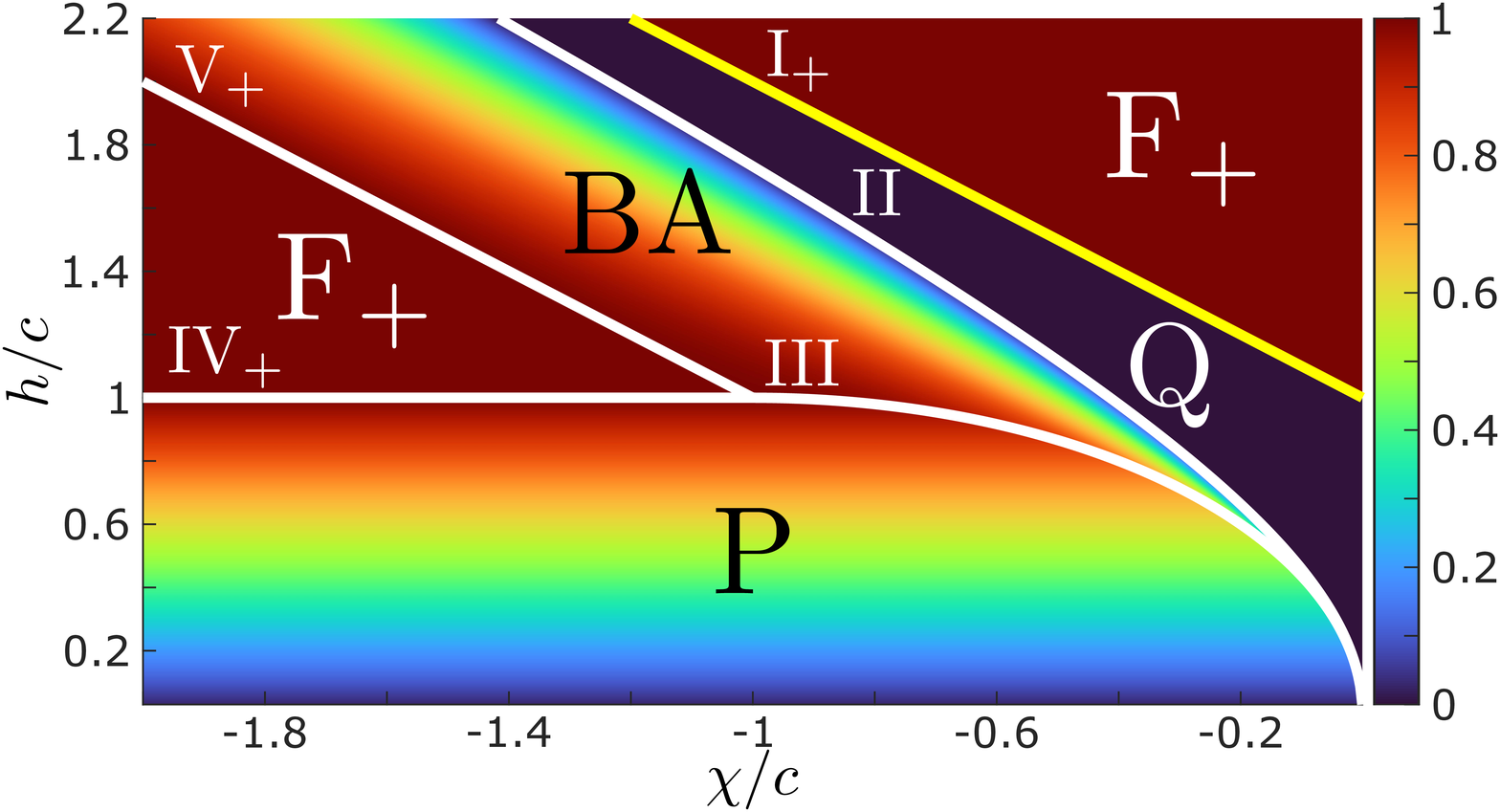}
    \label{fig:gg}    
    \end{minipage}
    \begin{minipage}[t]{0.49\linewidth}
    \includegraphics[width=77mm]{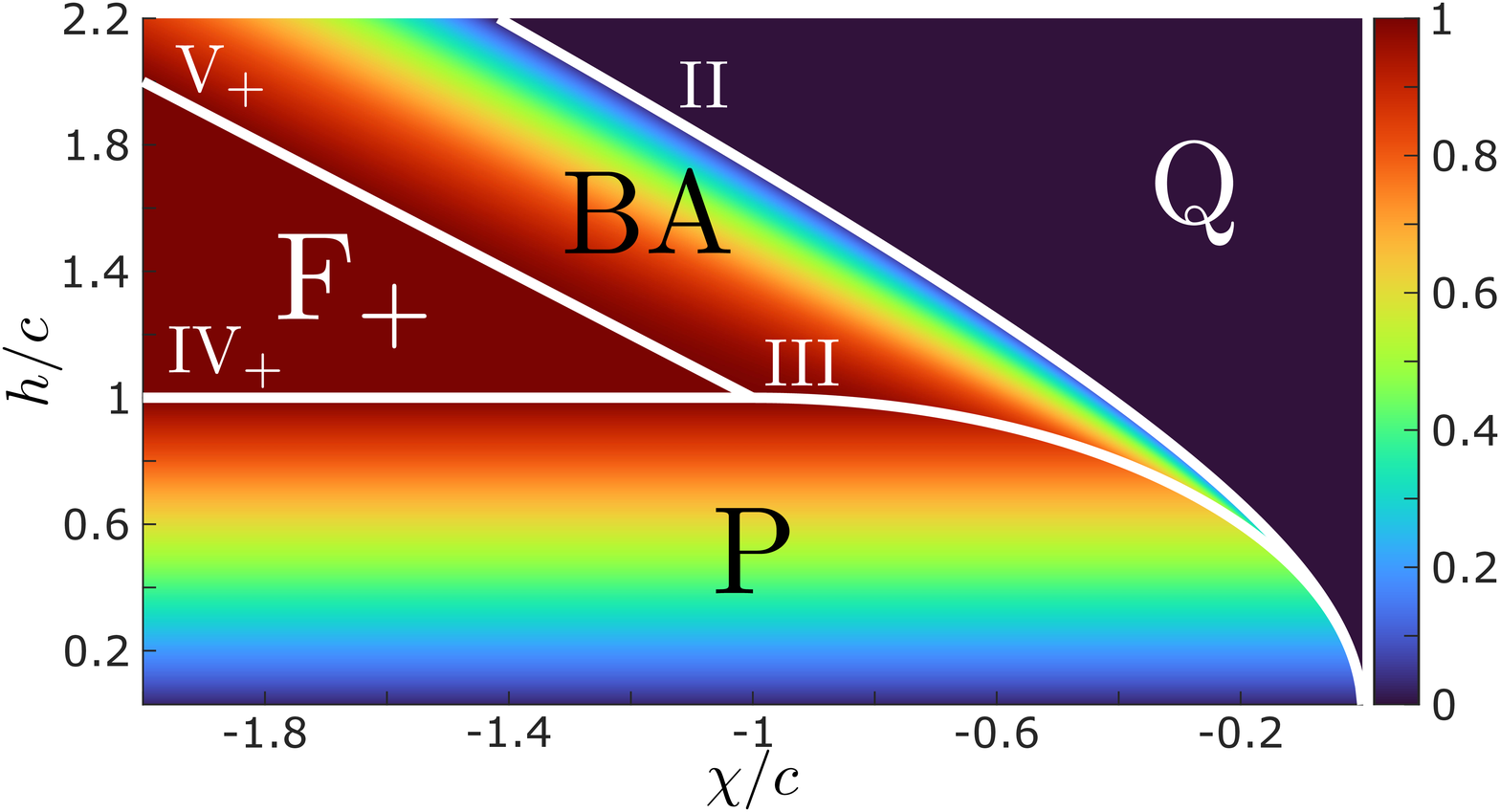}
    \end{minipage}
    \break
    \begin{minipage}[t]{0.49\linewidth}
    \includegraphics[width=77mm]{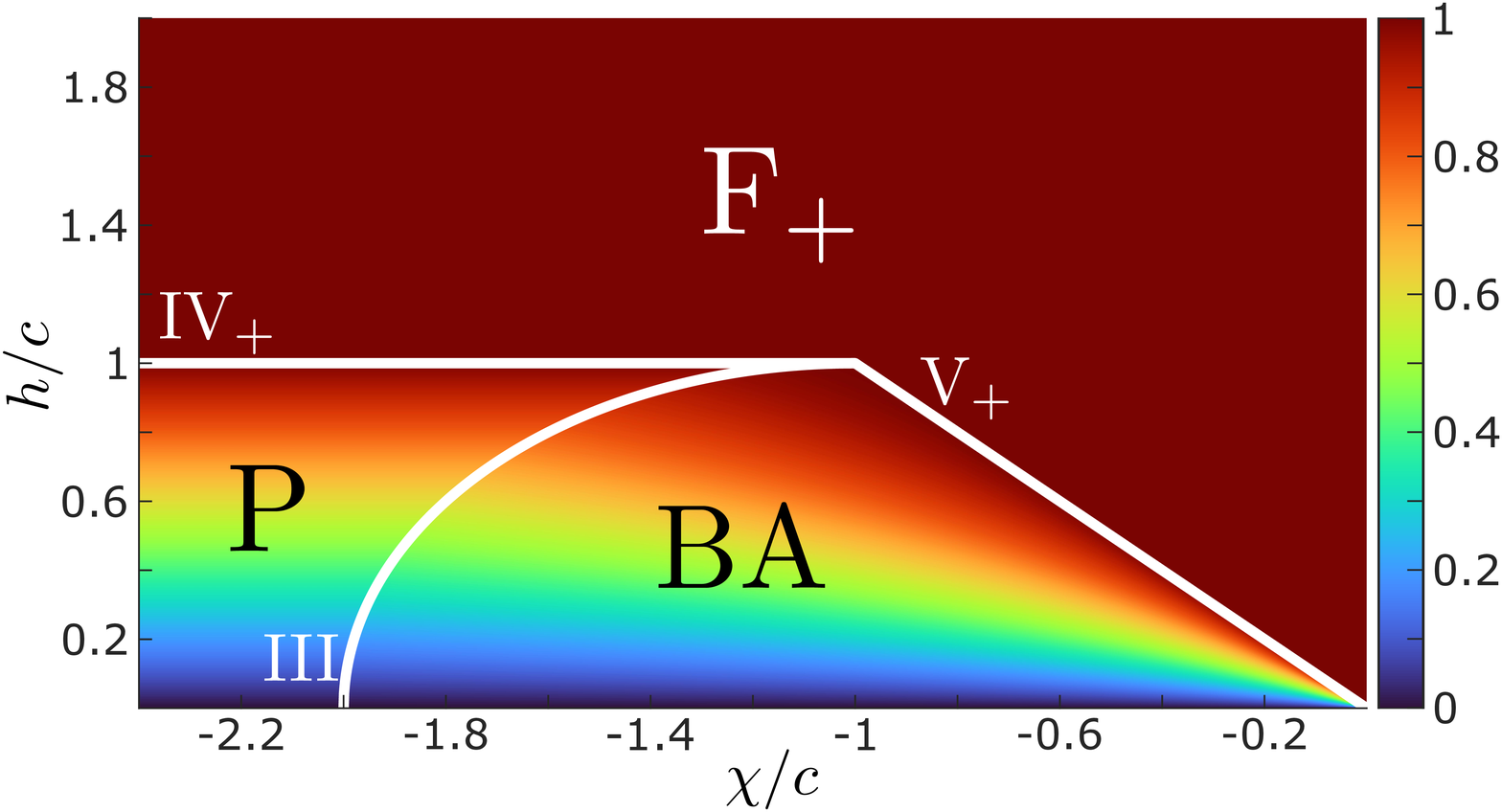}
    \end{minipage}
    \begin{minipage}[t]{0.49\linewidth}
    \includegraphics[width=77mm]{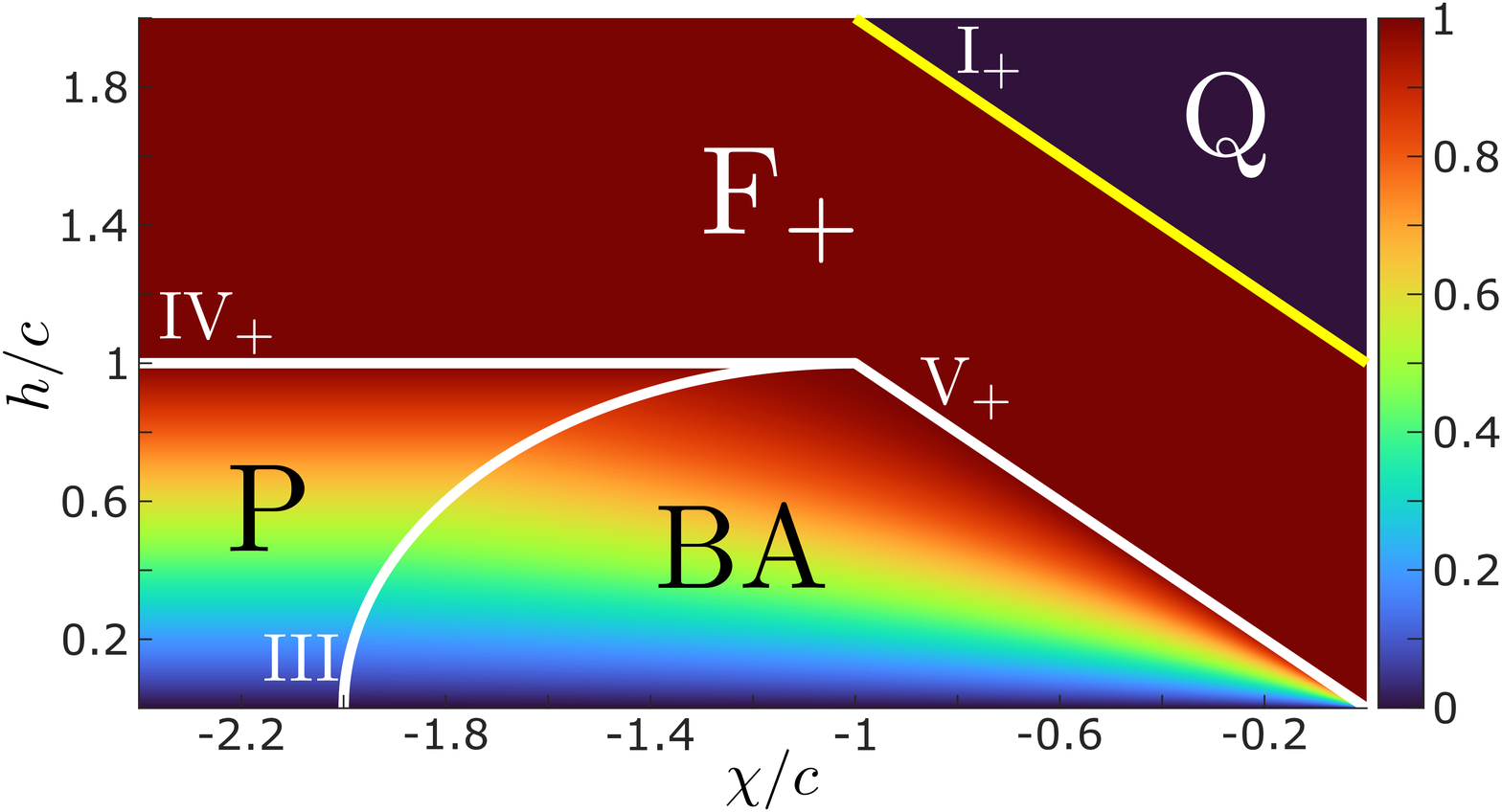}
    \end{minipage}
    \break
    \begin{minipage}[t]{0.49\linewidth}
    \includegraphics[width=77mm]{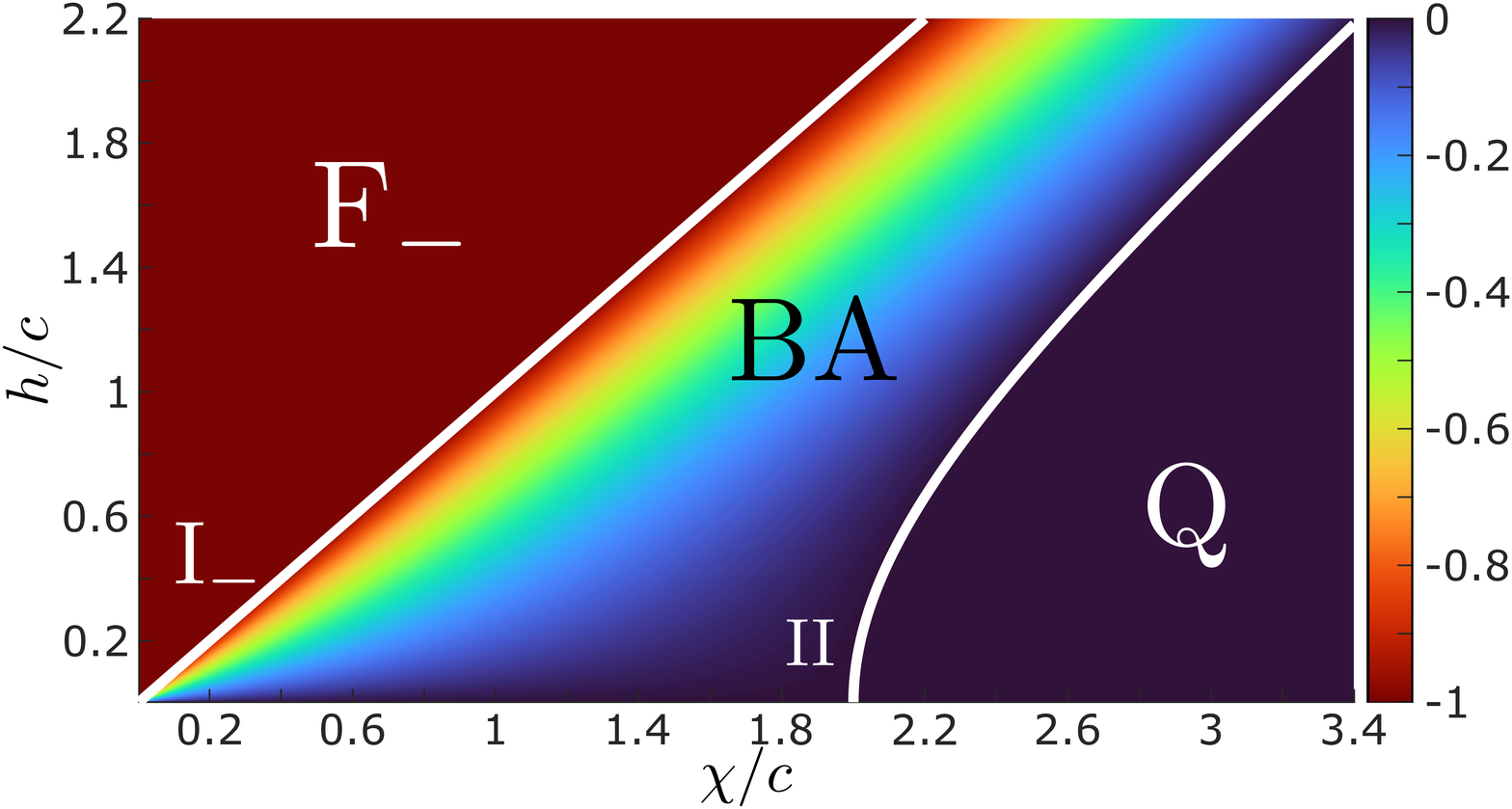}
    \end{minipage}
    \caption{Magnetic state diagrams of spin-1 BEC, showing magnetization $\langle S^{z}\rangle/n_{0} $ (by colour) versus dimensionless magnetic fields $\chi/c$ and $h/c$ for $c>0$ (left panels) and $c<0$ (right panels). Each row corresponds to certain regime of BA-state, see inequalities (\ref{ineq:cxl0xlh}), (\ref{ineq:cxl0xgh}), and (\ref{ineq:cxg0xgh}), respectively. The white and yellow lines denote the second- and first-order phase transtions, respectively. For details, see Table~\ref{tab:Separ}.
    }
    \label{fig:st_diag}
\end{figure}

Furthermore, each magnetic state is characterized by its own magnetization.
Therefore, it seems natural to depict all the examined states in a diagram in terms 
of magnetization $\langle S^{z}\rangle$ and applied external fields $h$ and $\chi$. 
Assuming that the system is prepared in the broken-axisymmetry state, we plot the mentioned magnetic state diagrams in Fig.~\ref{fig:st_diag} for all three regimes given by  inequalities \eqref{eq:ineq} and different signs of interaction parameter $c$. Note that two of them corresponding to (\ref{ineq:cxg0xgh}) coincide for both signs of $c$. The diagrams can be mirrored in the lower half plane, i.e., the region of negative values of $h/c$. All magnetic states are delimited by the separatrices identified with corresponding Roman numerals in Table~\ref{tab:Separ}. 
On the white lines, we are dealing with the second-order phase transition because both magnetization and quadrupole moments change continuously.  
On the contrary, we observe a jump of magnetization and quadrupole moment on the yellow lines, which correspond to the first-order phase transition.

Note that there is no way to realize the several regimes of BA-state on a single diagram. It is worth stressing that the possible succession of phase transitions in the diagrams is realized according to the above graphs, see Fig.~\ref{fig:graph}. Surely, we believe that the corresponding stability conditions are met for each magnetic state (see Table~\ref{tab:StabCond}).  

\section{Summary}
\label{Summary}
We have studied all possible magnetic states emerging in interacting spin-1 condensate under the general consideration of quadrupole degrees of freedom. 
The role of the latter is twofold: they provide the coupling between the quadrupole moment and applied magnetic field as well as generate the interatomic quadrupole-exchange interaction. 
In the context of ultracold atomic gas, the latter makes sense when considering the non-local effects of interatomic interaction. 
Indeed, for van der Waals forces ($V\propto r^{-6}$), the low-energy scattering can be safely described by the local ($\delta$-like) potential parameterized by the $s$-wave scattering length.    
Therefore, as we mentioned, the term $V_{K}$ (see Eqs.~(\ref{eq:InterHam}) and (\ref{eq:InterHamK})) is irrelevant one. Such description is valid for alkali atoms with a very small value of the magnetic moment. However, some atomic species, for example, Erbium (Er) \cite{Aikawa_PRL2012}, Dysprosium (Dy) \cite{Lu_PRL2011} and their isotopes have large magnetic moment of the order of $5\--10 \mu_{\rm B}$ and their interaction is specified by the long-range dipole force 
$$
V_{dd}(r,\theta)=\frac{\mu_{0}\mu_{M}^{2}}{4\pi}\frac{1-3\cos^{2}\theta}{r^{3}},
$$
where $\mu_{M}$ is the permanent magnetic dipole moment and $\mu_{0}$ is the vacuum permittivity.
Here we assume that the dipole moments are oriented along the same direction, fixed by an external field ($\theta$ is the angle between the dipoles). It is useful to characterize the intensity of interaction by the following effective range \cite{PitStr2016}: 
\begin{equation*}
r_{dd}=\frac{\mu_{0}\mu_{M}^{2}m}{12\pi\hbar^{2}}.
\end{equation*}
Thus, the ratio $\varepsilon_{dd}=r_{dd}/a$ allows one to compare the intensities of the dipole-dipole interaction and zero-range interaction specified by the scattering length $a$. The effective ranges, $r_{dd}$, for ${}^{168}$Er and ${}^{164}$Dy are $3.7$\,nm and $7$\,nm, respectively \cite{PitStr2016}. The $s$-wave scattering lengths arising from the short-range effects are of the order $100a_{0}$ ($\sim5.3\,nm$) for both atomic species and their isotopes, where $a_{0}$ is the Bohr radius (see Ref.~\cite{Ferlaino_PRA2022} for Erbium and Ref.~\cite{Lev_PRA2015} for Dysprosium). Therefore, $\varepsilon_{dd}\sim 1$ in both cases. This indicates the relative importance of long-range (non-local) interaction in such atomic gases, and, consequently, the quadrupole exchange interaction term $V_{K}$ given by Eq.~(\ref{eq:InterHamK}) become relevant since the quantities $U(0)$, $J(0)$, and $K(0)$ can be of the same order of magnitude.

In contrast to ferromagnetic, quadrupolar, and paramagnetic states, the coupling of quadrupole moment with external field is the necessary condition for the broken-axisymmetry state to exist. This state has a number of unique properties: all components of the state vector (three spin projections) are occupied, magnetization vector is not parallel to the applied magnetic field, and expectation value of the quadrupole operator has no zero matrix elements indicating a significant magnetic anisotropy. Therefore, one can claim that the existence of BA-state is not caused by the quadratic Zeeman effect itself, but the related magnetic anisotropy. Obviously, the latter can be produced in many experimental ways \cite{Ketterle1998,Dalibard2012,Gerbier2006,Leslie2009,Santos2007}. The BA-state in regime (\ref{ineq:cxg0xgh}) was first studied experimentally \cite{Ketterle1998} and theoretically \cite{UedaPRA2007}  (see the third row in Fig.~\ref{fig:st_diag}). The phase transition from BA-state to paramagnetic one across separatrix III (see Table \ref{tab:Separ} and the second row in Fig.~\ref{fig:st_diag}) was experimentally examined in Ref.~\cite{Dalibard2012}. In the presence of additional quadrupole-exchange interaction, we analyzed all the regimes of the BA-state to exist. 
We plotted the magnetic state diagrams to visualize the regimes and corresponding phase transitions to other magnetic states.

Finally, since we manage with two quantities $v$ and $c$ associated with two coupling constants (see Eqs.~\eqref{eq:local}, \eqref{eq:Interparam}, Tables 3 and 4) there is no qualitative difference between the magnetic state diagrams for local and non-local interactions. Nevertheless, we have obtained the additional magnetic state diagrams at other regimes of a magnetic field that have not been previously studied either theoretically or experimentally. This opens up the possibility of their experimental confirmation even in the case of local interaction. In addition, the effects associated with non-locality of interaction manifest themselves in the spectrum of quasiparticle excitations (see Eqs.~\eqref{eq:spec1}-\eqref{eq:spec2}). Therefore, they can be revealed in the experimental study. To the best of our knowledge, such experiments have not yet been carried out and we hope them to be possible in the future.

\section*{Acknowledgements}
The authors are grateful to Andrii Sotnikov for useful discussions. We thank an anonymous reviewer for his helpful guidance, which allowed us to improve the quality of the paper. The authors acknowledge support from the National Research Foundation of Ukraine, Grant No. 0120U104963, Ministry of Education and Science of Ukraine, Research Grant No. 0122U001575, and National Academy of Sciences of Ukraine, Project No. 0121U108722.

\begin{appendix}
\section{Gell-Mann generators of the SU(3) group}\label{app:Gell}
\setcounter{section}{1}
The Gell-Mann matrices representing the generators of the SU(3) group are given by
\begin{gather}
\lambda^{1}=\left(
  \begin{array}{ccc}
    0 & 1 & 0 \\ 
    1 & 0 & 0 \\ 
    0 & 0 & 0 \\
  \end{array}
\right), \quad
\lambda^{2}=\left(
  \begin{array}{ccc}
    0 & -i & 0 \\
    i & 0 & 0 \\
    0 & 0 & 0 \\
  \end{array}
\right), \quad
\lambda^{3}=\left(
  \begin{array}{ccc}
    1 & 0 & 0 \\ 
    0 & -1 & 0 \\ 
    0 & 0 & 0 \\
  \end{array}
\right), \nonumber \\
\lambda^{4}=\left(
  \begin{array}{ccc}
    0 & 0 & 1 \\
    0 & 0 & 0 \\
    1 & 0 & 0 \\
  \end{array}
\right), \quad
\lambda^{5}=\left(
  \begin{array}{ccc}
    0 & 0 & -i \\ 
    0 & 0 & 0 \\ 
    i & 0 & 0 \\
  \end{array}
\right), \quad
\lambda^{6}=\left(
  \begin{array}{ccc}
    0 & 0 & 0 \\
    0 & 0 & 1 \\
    0 & 1 & 0 \\
  \end{array}
\right), \nonumber \\
\lambda^{7}=\left(
  \begin{array}{ccc}
    0 & 0 & 0 \\
    0 & 0 & -i \\ 
    0 & i & 0 \\
  \end{array}
\right), \quad
\lambda^{8}=\frac{1}{\sqrt{3}}\left(
  \begin{array}{ccc}
    1 & 0 & 0 \\
    0 & 1 & 0 \\
    0 & 0 & -2 \\
  \end{array}
\right).\label{eq:AGellMann}
\end{gather}
The Hermitian and traceless operators $\lambda^{a}$, have the following property:
\begin{equation}\label{eq:A2}
{\rm Tr}\lambda^{a}\lambda^{b}=2\delta^{ab}
\end{equation}
and meet the following commutation relations:
\begin{equation}\label{eq:A3}
[\lambda^{a},\lambda^{b}]=2if^{abc}\lambda^{c},
\end{equation}
where $f^{abc}$ are the structure constants of the SU(3) group. From Eq.~(\ref{eq:A2}), one obtains
$$
f^{abc}=-\frac{i}4{\rm Tr}\,\lambda^{c}[\lambda^{a},\lambda^{b}],
$$
wherefrom
\begin{equation} \label{eq:A4}
f^{abc}=-f^{bac}=f^{bca}.
\end{equation}
The structure constants $f^{abc}$ have the following numerical values:
\begin{equation}
\label{eq:A5}
f^{123}=1, \quad f^{147}=-f^{156}=f^{246}=f^{257}=f^{345}=-f^{367}=\frac12, \quad f^{456}=f^{678}=\frac{\sqrt{3}}{2}.
\end{equation}
All other numerical values of $f^{abc}$ not related to the indicated above by permutation are
zero. The anticommutator of the Gell-Mann matrices, as well as the commutator, is linear in $\lambda_{a}$:
\begin{equation} \label{eq:AAntiCom}
\{\lambda^{a},\lambda^{b}\}=\frac43\delta^{ab}+2d^{abc}\lambda^{c},
\end{equation}
where the coefficients $d^{abc}$, symmetric over all indices, are given by
$$
d^{abc}=\frac14{\rm Tr}\,\lambda^{c}\{\lambda^{a},\lambda^{b}\}.
$$
The following values of $d^{abc}$ are different from zero:
\begin{gather}
d^{118}=d^{228}=d^{338}=-d^{888}=\frac{1}{\sqrt{3}}, \nonumber \\
d^{146}=d^{157}=d^{256}=d^{344}=d^{355}=-d^{247}=-d^{366}=-d^{377}=\frac12, \nonumber \\
d^{448}=d^{558}=d^{668}=d^{778}=-\frac{1}{2\sqrt{3}}. \label{eq:A8}
\end{gather}

\section{Solving equation for the vector order parameter}
\label{ap:sol}
In explicit form the coupled equations~\eqref{eq:MinCond} reads
\begin{subequations}
\label{eq:explict_min}
\begin{eqnarray}
	\label{eq:explict_min:a}
	a
	\zeta_x
	+
	c
	\left(
	\zeta_x^2
	+
	\zeta_y^2
	+
	\zeta_z^2
	\right)
	\zeta_x^*
	-
	i
	h
	\zeta_y
	=
	0,
	\\\label{eq:explict_min:b}
	a
	\zeta_y
	+
	c
	\left(
	\zeta_x^2
	+
	\zeta_y^2
	+
	\zeta_z^2
	\right)
	\zeta_y^*
	+
	i
	h
	\zeta_x
	=
	0,
	\\\label{eq:explict_min:c}
	(a-\chi)
	\zeta_z
	+
	c
	\left(
	\zeta_x^2
	+
	\zeta_y^2
	+
	\zeta_z^2
	\right)
	\zeta_z^*
	=
	0,
\end{eqnarray}
\end{subequations}
where 
$
	a=		
	\tilde{\mu}
	+
	\chi
	-
	v
	-
	c
$ and
$	
	c=
	n_0
	\left(
		J(0)
		-
		K(0)
	\right)
$.

Note that $\boldsymbol{\zeta}=(\zeta_{x},\zeta_{y},\zeta_{z})^{T}$ represents the order parameter in the Cartesian (vector) basis, defined as $S^{i}|i\rangle=0$, $i=x,y,z$. In this basis, the problem of inetarcting spin-1 atoms is formulated in an elegant way because the corresponding spin matrices $(S^{i})_{kl}=-i\varepsilon_{ikl}$ form a subalgebra for the Gell-Mann generators of the SU(3) group. Further, it is more intuitive and convenient to use the canonical basis $|m\rangle$, which is defined as the eigenstate of $S^{z}$, $S^{z}|m\rangle=m|m\rangle$, where $m=-1,0,1$ are the spin projections. The spinor order parameters in these two bases are related by
\begin{equation}
\label{eq:bastrans}
\boldsymbol{\zeta}'= R\boldsymbol{\zeta}, \quad R=\frac{1}{\sqrt{2}}\left(
  \begin{array}{ccc}
    -1 & i & 0 \\ 
    0 & 0 &  \sqrt{2}\\ 
    1 & i & 0 \\
  \end{array}
\right), 
\quad
R^{\dagger}R=1,
\end{equation}
where $\boldsymbol{\zeta}^{\prime}=(\zeta_{+},\zeta_{0},\zeta_{-})^{T}$ is the state vector in the canonical basis. For ferromagnetic (F), quadrupolar (Q), and paramagnetic (P) states, the corresponding spinor order parameters are indicated in the Table~\ref{tab:3states}, while for the broken-axisymmetry state it is given by Eqs.~(\ref{eq:BAChem}) and (\ref{eq:BAState}).

Employing the transformation \eqref{eq:bastrans} with subsequent algebraic recasting, we can write the system \eqref{eq:explict_min} through the state vector components in the canonical basis:
\begin{subequations}
\label{eq:state_mp}
\begin{alignat}{2}
\label{eq:state_mp:a}
	(a-h)
	\zeta_{-}
	+
	c
	\left(
	2
	\zeta_{-}
	\zeta_{+}
	-
	\zeta_0^2
	\right)
	\zeta_{+}^*
	&&=
	0,
	\\
\label{eq:state_mp:b}
    -(a-\chi)
	\zeta_0
	+
	c
	\left(
	2
	\zeta_{-}
	\zeta_{+}
	-
	\zeta_0^2
	\right)
	\zeta_0^*
	&&=
	0,
	\\	
	\label{eq:state_mp:c}
	(a+h)
	\zeta_{+}
	+
	c
	\left(
	2
	\zeta_{-}
	\zeta_{+}
	-
	\zeta_0^2
	\right)
	\zeta_{-}^*
	&&=
	0
	.
\end{alignat}
\end{subequations}

Let us briefly describe the algorithm for solving 
the coupled equations (\ref{eq:state_mp}).
We begin with the cases  when one of the state vector components $\zeta_i$ with $i=\{+,0,-\}$ is assumed to be equal to zero. 

First, if $\zeta_0=0$, then we obtain the reduced system for two 
sought variables with the following solutions: 
\begin{eqnarray*}
    a=\pm h,
\qquad
    &&
    \zeta_0=\zeta_{\pm}=0,\quad\zeta_{\mp}=1
,\\
    a=-c,
\qquad
    &&
	\zeta_0=0
	,\quad
	\left|
		\zeta_{\pm}
	\right|
	^2
	=
	\frac12
	\left(
		1
		\pm
		\frac{h}{c}
	\right).
\end{eqnarray*}
They correspond to ferromagnetic (F) and paramagnetic (P) states (see Table~\ref{tab:3states}), respectively. Second, if we put $\zeta_{\pm}=0$, the solution, respectively, is $\zeta_{\mp}=0$ and $\zeta_0=1$ with $a=\chi-c$.
Obviously, this solution corresponds to the quadrupolar (Q) state (see Table~\ref{tab:3states}). 

Some more steps are required to obtain the most nontrivial solution with all non-zero components of the state vector, $\zeta_i\neq 0$ with $i=\{+,0,-\}$:  
\begin{itemize}
    \item to multiply each equation from (\ref{eq:state_mp}) on $\zeta_{+}$, $\zeta_{0}$, and $\zeta_{-}$, respectively;
    \item to sum the result equations and to employ the normalization condition;
    \item to express the combination $2\zeta_{+}\zeta_{-}$ through $\zeta_{0}^2$,
    \begin{equation}\label{eq:zpzm}
    2\zeta_{+}\zeta_{-}=\frac{a+c-\chi}{a+c}\zeta_{0}^{2};
    \end{equation} 
    \item to substitute this relation into every equation obtained on the first step;
    \item to find the modulus of each state vector component.
\end{itemize}
As a result, we have
\begin{equation*}
    |\zeta_{+}|=\sqrt{
    \frac{
        (a-h)(a+c-\chi)
    }{
        2c\chi
    }
    }
    ,\quad
    |\zeta_{0}|=\sqrt{
    \frac{
        (\chi-a)(a+c)
    }{
        c\chi
    }
    }
    ,\quad
    |\zeta_{-}|=\sqrt{
    \frac{
        (a+h)(a+c-\chi)
    }{
        2c\chi
    }
    }
    .
\end{equation*}
Unfortunately, this procedure does not yield a relation for the chemical potential $\tilde{\mu}$. For this purpose, we should make several additional steps:
\begin{itemize}
    \item to replace $\zeta_{0}^2$ in Eq.~(\ref{eq:state_mp:b})  by means of Eq.~(\ref{eq:zpzm}); 
    \item to replace each state vector component with product  $\zeta_{i}=|\zeta_{i}|e^{i\phi_i}$.
\end{itemize}
As a result, we get
\begin{equation}\label{eq:ForPhase}
    (a-\chi)e^{i\phi_0}
    =
    \frac{
        2c\chi
    }{
        a+c-\chi
    }|\zeta_{+}||\zeta_{-}|e^{i(\phi_++\phi_--\phi_0)}
    .
\end{equation}
Therefore, we obtain the constraint on phases (\ref{eq:BAPhases})
and equation for determining $a$ (or chemical potential $\tilde{\mu}$). The latter has only one solution,
\begin{equation*}
    a=\frac{h^2+\chi^2}{2\chi}
    \quad
    \Rightarrow
    \quad
    \tilde{\mu}
    =
    \frac{h^{2}-\chi^{2}}{2\chi}+v+c
    .
\end{equation*}

\end{appendix}
\bibliographystyle{iopart-num}
\bibliography{main}

\end{document}